\documentclass[
 reprint,
 showpacs, preprintnumbers,
 amsmath, amssymb,
 aps,
 pra,
 floatfix,
 superscriptaddress
]{revtex4-1}

\usepackage{graphicx}
\usepackage{tikz}
\usepackage{xcolor}
\usetikzlibrary{arrows}

\makeatletter
\newcommand{\colorcaption}[2][]{%
  \begingroup%
  \renewcommand{\@caption@fignum@sep}{ (Color online). }%
  \caption[#1]{#2}%
  \endgroup%
}
\makeatother

\makeatletter
\def\maketag@@@#1{\hbox{\m@th\normalfont\normalsize#1}}
\makeatother

\usepackage{dcolumn}
\usepackage{bm}
\usepackage{bbm}
\usepackage{upgreek}
\usepackage{hyperref}
\usepackage{blindtext}
\usepackage{hyphsubst}

\usepackage{changes}

\usepackage{dsfont}

\usepackage{newtxtext,newtxmath}

\definecolor{darkgreen}{RGB}{0 100 0}

\begin{document}

	\preprint{}


	\title{Superconducting triplet pairings and anisotropic magnetoresistance effects \\ in ferromagnet/superconductor/ferromagnet double-barrier junctions}
	
	\author{Andreas Costa}%
	\email[Corresponding author: ]{andreas.costa@physik.uni-regensburg.de}
 	\affiliation{Institute for Theoretical Physics, University of Regensburg, 93040 Regensburg, Germany}
 	
 	\author{Jaroslav Fabian}%
 	\affiliation{Institute for Theoretical Physics, University of Regensburg, 93040 Regensburg, Germany}

 	\date{\today}
    
    
    \begin{abstract}
    	Ferromagnetic spin~valves offer the key building~blocks to integrate giant- and tunneling-magnetoresistance effects into spintronics~devices. 
    	Starting from a generalized Blonder--Tinkham--Klapwijk~approach, we theoretically investigate the impact of interfacial Rashba and Dresselhaus spin-orbit~couplings on the tunneling~conductance, and thereby the magnetoresistance~characteristics, of ferromagnet/superconductor/ferromagnet spin-valve~junctions embedding thin superconducting spacers between the either parallel or antiparallel magnetized ferromagnets. 
    	We focus on the unique interplay between usual electron~tunnelings---that fully determine the magnetoresistance in the normal-conducting state---and the peculiar Andreev~reflections in the superconducting state. 
    	In the presence of interfacial spin-orbit~couplings, special attention needs to be paid to the spin-flip~(``\emph{unconventional}'') Andreev-reflection~process that is expected to induce superconducting triplet~correlations in proximitized regions. 
    	As a transport~signature of these triplet~pairings, we detect conductance double~peaks around the singlet-gap~energy, reflecting the competition between the singlet and an additionally emerging triplet gap; the latter is an effective superconducting gap that can be ascribed to the formation of triplet Cooper~pairs through interfacial spin-flip~scatterings~(i.e., to the generation of an effective triplet-pairing term in the order parameter). 
    	We thoroughly analyze the Andreev~reflections' role in connection with superconducting magnetoresistance~phenomena, and eventually unravel huge conductance and magnetoresistance magnetoanisotropies---easily exceeding their normal-state counterparts by several orders of magnitude---as another experimentally accessible fingerprint of unconventional Andreev~reflections. 
    	Our results provide an important contribution to establish superconducting magnetic spin~valves as an essential ingredient for future superconducting-spintronics concepts. 
    \end{abstract}
    
    
    \maketitle
    

    \section{Introduction   \label{Sec_Introduction}}

    The tunneling-magnetoresistance~(TMR)~effect~\cite{Julliere1975, Moodera1995}, occurring when switching the \emph{relative} magnetizations of  ferromagnet/insulator/ferromagnet~(F/I/F) spin~valves' metallic layers, is one of the most spectacular spintronics~phenomena~\cite{Fabian2004,Fabian2007}, especially
    considering its technological applications in computers~\cite{Fert2008,Hirota2002,Parkin2002,Slaughter2003,Bhatti2017}.
    Numerous proposals to engineer next-level quantum~computers have been put forward within recent years~\cite{Ioffe1999,Mooij1999,Blatter2001,Ustinov2003,Yamashita2005,Feofanov2010,Khabipov2010,Devoret2013}, and might come along with a so far unimaginable boost of computing~performance. 
    Owing to its great advantages~\cite{Eschrig2011,Linder2015} when combining quantum~coherence, which belongs to the most fundamental ingredients for quantum~computing, with dissipationless charge and long-lived spin~transport, most of the aforementioned concepts exploit superconductivity.

    Among the systems attracting the most considerable interest are superconducting magnetic tunnel~junctions, in which the competition between the two nominally strongly antagonistically acting superconducting and ferromagnetic phases offers a versatile playground to study novel physical characteristics. 
    While early works focused, e.g., on the conductance of ferromagnet/superconductor point~contacts~\cite{Soulen1998,Soulen1999}, thereby demonstrating that Andreev~reflections impact transport in a unique way from which the ferromagnet's spin~polarization can be experimentally extracted, more intricate junction setups are being explored nowadays. 
    Magnetic Josephson~junctions are particularly promising candidates to investigate unprecedented transport~anomalies, covering current-reversing $ 0 $-$ \pi $~transitions~\cite{Bulaevskii1977a,*Bulaevskii1977b,Ryazanov2001} that could form the elementary two-level system for quantum~computing, substantially enhanced current~magnetoanisotropies~\cite{Hoegl2015,*Hoegl2015a,Jacobsen2016,Costa2017,Martinez2020}, the potential appearance of Majorana~states~\cite{Nilsson2008,Duckheim2011,Lee2012a,Nadj-Perge2014,Dumitrescu2015,Pawlak2016,Ruby2017,Livanas2019,Manna2020}, as well as the possibility to efficiently generate long-range spin-polarized triplet-Cooper~pair supercurrents~\cite{Keizer2006,Eschrig2011}. 
    Such triplet~pairings are typically induced in $ s $-wave superconductors proximitized by (strongly spin-polarized) ferromagnets either in the presence of noncollinearly magnetized interfacial domains~\cite{Bergeret2001b,Bergeret2001,Bergeret2001a,Eschrig2003,Houzet2007,Eschrig2008,Grein2009,Khaire2010,Robinson2010a,Robinson2010,Banerjee2014,Diesch2018} or spin-orbit~coupling~(SOC)~\cite{Bergeret2013,Bergeret2014,Costa2017}.

    The normal-state TMR~effect's superconducting counterpart was already investigated in theoretical~\cite{DeGennes1966a} and experimental~\cite{Hauser1969,Li2013} works carried out on ferromagnet/superconductor/ferromagnet~(F/S/F) spin~valves, in which a thin superconducting spacer couples the parallel or antiparallel magnetized ferromagnetic electrodes. 
    Remarkably, flipping the magnetizations from antiparallel to parallel may decrease the superconductor's critical temperature. 
    Close to the critical temperature, the magnetization flipping can thus switch off superconductivity, resulting in an infinitely large TMR~ratio. 
    While the normal-state TMR was fully explained by spin-polarized electron~tunnelings~\cite{Julliere1975}, 
    Andreev~reflections~\cite{Andreev1964,*Andreev1964alt} can as well strongly influence electrical transport in superconducting junctions and modify the TMR~characteristics there~\cite{Bozovic2002,*Bozovic2005}.

    In this paper, we investigate the influence of SOC on the transport~properties of F/S/F~junctions. We assume thin semiconducting tunneling~barriers between the ferromagnetic and superconducting regions with interfacial (Bychkov--)Rashba~\cite{Bychkov1984b,*Bychkov1984c} and Dresselhaus~\cite{Dresselhaus1955} SOCs. Such couplings occur 
    in the presence of semiconductors such as InAs or InSb, whose
    (001)~interfaces have $ C_{2v} $~symmetry~\cite{Fabian2007}.
    As mentioned above, these SOCs induce effective superconducting triplet~pairings close to the junction~interfaces, microscopically mediated by spin-flip~(``\emph{unconventional}'') Andreev~reflections, just as the usual spin-conserving~(``conventional'') Andreev~reflections bring singlet superconducting order into proximitized junction~regions. 
    The unconventional Andreev~reflections are extremely sensitive to changing \emph{absolute} magnetization~orientations, and are thus at the heart of the huge magnetoanisotropies in superconducting magnetic junctions~\cite{Hoegl2015,*Hoegl2015a,Jacobsen2016,Costa2017,Martinez2020}.

    Generalizing the well-established Blonder--Tinkham--Klapwijk~model~\cite{Blonder1982} to spin-valve~junctions~\cite{Moen2017,Moen2018a,Moen2018}, we evaluate the junctions' \emph{zero-temperature} tunneling~conductance and elaborate on transport~ramifications of unconventional Andreev~reflections, predicting the formation of conductance double~peaks close to the singlet-gap~energy as a consequence of the effectively induced nonzero triplet gap. 
    After demonstrating the expected huge magnetosensitivity of the double-peak~conductance~structure, we compute typical (T)MR~ratios~\footnote{Strictly speaking, the term \emph{tunneling~magnetoresistance~(TMR)} is only used for junctions in the tunneling~limit, i.e., if strong interfacial tunneling~barriers were present. As we consider weak barriers to elucidate the physics related to Andreev~reflections, we shall rather call the effect simply \emph{magnetoresistance~(MR)}. } and identify marked MR~magnetoanisotropies, which provide another clear fingerprint of unconventional Andreev~reflections. 
    Our predictions should help experiments in disentangling triplet and singlet superconducting proximity pairings in F/S/F~spin~valves' tunneling~conductance.

    We structure the paper in the following way. 
    In~Sec.~\ref{Sec_Model}, we formulate our theoretical model to describe electrical transport through the considered F/S/F~junctions. 
    The general conductance~features and double~peaks as signatures of triplet~pairing are thoroughly analyzed in~Sec.~\ref{Sec_General}, before we briefly comment on the conductance's magnetic~tunability~(magnetoanisotropy) in~Sec.~\ref{Sec_MAAR}. 
    The MR and its magnetoanisotropy are discussed in~Secs.~\ref{Sec_TMR} and \ref{Sec_ATMR}, respectively, while we conclude our main findings in~Sec.~\ref{Sec_Conclusions}.

    \section{Theoretical model  \label{Sec_Model}}

\begin{figure}
        \centering
        \includegraphics[width=0.40\textwidth]{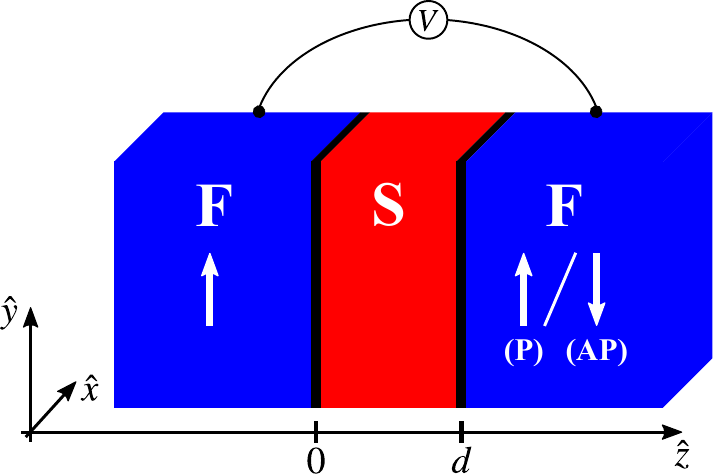}
        \caption{Sketch of the considered F/S/F~junction, using $ C_{2v} $~crystallographic~orientations $ \hat{x} \parallel [110] $, $ \hat{y} \parallel [\overline{1}10] $, and $ \hat{z} \parallel [001] $. 
            The junction's ferromagnetic electrodes~(F; blue) and the superconducting link~region of thickness~$ d $~(S; red) are separated through ultrathin semiconducting tunneling~barriers~(black), additionally introducing interfacial SOC. 
            The ferromagnets' magnetizations can be aligned either parallel~(P) or antiparallel~(AP) with respect to each other, as illustrated by the white arrows, while $ V $ denotes the applied bias~voltage. 
        }
        \label{Fig_Junction}
    \end{figure}

    We consider a biased ballistic F/S/F~junction grown along the $ \hat{z} \parallel [001] $ crystallographic direction, in which the two semi-infinite ferromagnetic electrodes~(F) are separated from the superconducting~link~(S) by ultrathin semiconducting tunneling~barriers~(see~Fig.~\ref{Fig_Junction}) with $ C_{2v} $~symmetry. 

    We model our system by means of its stationary Bogoliubov--de~Gennes Hamiltonian~\cite{DeGennes1989}
    \begin{equation}
        \hat{\mathcal{H}}_\mathrm{BdG} = \left[ \begin{matrix} \hat{\mathcal{H}}_\mathrm{e} & \hat{\Delta}_\mathrm{S}(z) \\ \hat{\Delta}_\mathrm{S}^\dagger (z) & \hat{\mathcal{H}}_\mathrm{h} \end{matrix} \right] ,
    \end{equation}
    where 
    \begin{align}
        \hat{\mathcal{H}}_\mathrm{e} &= \left( - \frac{\hbar^2}{2m} \boldsymbol{\nabla}^2 - \mu \right) \hat{\sigma}_0 \nonumber \\
        &\hspace{25 pt} - \frac{\Delta_\mathrm{XC}}{2} (\hat{\mathbf{m}}_\mathbf{1} \cdot \hat{\boldsymbol{\sigma}}) \Theta (-z) \nonumber \\
        &\hspace{25 pt} - \frac{\Delta_\mathrm{XC}}{2} (\hat{\mathbf{m}}_\mathbf{2} \cdot \hat{\boldsymbol{\sigma}}) \Theta (z-d) \nonumber \\
        &\hspace{50 pt} + ( V_\mathrm{L} d_\mathrm{L} \hat{\sigma}_0 + \boldsymbol{\Omega}_\mathbf{L} \cdot \hat{\boldsymbol{\sigma}} ) \delta(z) \nonumber \\
        &\hspace{50 pt} + ( V_\mathrm{R} d_\mathrm{R} \hat{\sigma}_0 + \boldsymbol{\Omega}_\mathbf{R} \cdot \hat{\boldsymbol{\sigma}} ) \delta (z-d)
        \label{Eq_SingleHamiltonian}
    \end{align}
    represents the Hamiltonian of single electrons and $ \hat{\mathcal{H}}_\mathrm{h} = - \hat{\sigma}_y \hat{\mathcal{H}}_\mathrm{e}^{*} \hat{\sigma}_y $ its holelike counterpart~($ \hat{\sigma}_0 $ and $ \hat{\sigma}_i $ indicate the $ 2 \times 2 $~identity and the $ i $th Pauli~matrix; $ \hat{\boldsymbol{\sigma}} = [ \hat{\sigma}_x , \, \hat{\sigma}_y , \, \hat{\sigma}_z ]^\top $ is the vector of Pauli~matrices). 
    Both ferromagnetic electrodes are described within the Stoner~model with the \emph{same exchange~energy}~$ \Delta_\mathrm{XC} $, and the, in general,  \emph{different in-plane magnetization~directions}~$ \hat{\mathbf{m}}_\mathbf{1} = [ \cos \Phi_1 , \, \sin \Phi_1 , \, 0 ]^\top $ in the left and $ \hat{\mathbf{m}}_\mathbf{2} = [ \cos \Phi_2 , \, \sin \Phi_2 , \, 0 ]^\top $ in the right F; the angles $ \Phi_1 $ and $ \Phi_2 $ are thereby measured with respect to the $ \hat{x} $~reference~axis, which is taken to be the principal-symmetry [110] crystallographic axis. 
    
    Following earlier studies~\cite{DeJong1995,Zutic1999,Zutic2000,Costa2017,Costa2018,Costa2019,Costa2020}, the ultrathin semiconducting interface~layers are included into our model as deltalike barriers with heights~(widths)~$ V_\mathrm{L} $~($ d_\mathrm{L} $) at the left and $ V_\mathrm{R} $~($ d_\mathrm{R} $) at the right interface, respectively. 
    Their SOCs enter through the effective spin-orbit~fields~$ \mathbf{\Omega}_\mathbf{L} = [ ( \alpha_\mathrm{L} - \beta_\mathrm{L} ) k_y , \, -( \alpha_\mathrm{L} + \beta_\mathrm{L} ) k_x , \, 0] $ and $ \mathbf{\Omega}_\mathbf{R} = - [ ( \alpha_\mathrm{R} - \beta_\mathrm{R} ) k_y , \, -( \alpha_\mathrm{R} + \beta_\mathrm{R} ) k_x , \, 0] $, where the terms scaling with the effective SOC~strength~$ \alpha_\mathrm{L} $~($ \alpha_\mathrm{R} $) account for the Rashba~SOC at the left~(right) interface and the remaining ones for linearized Dresselhaus~SOC with the effective strengths~$ \beta_\mathrm{L} $~($ \beta_\mathrm{R} $). Note that we define the sign of $ \alpha_\mathrm{R}$ opposite to that of 
    $ \alpha_\mathrm{L}$, reflecting the fact that Rashba coupling arises from interfacial hybridization.

    Inside the superconducting link, the $ s $-wave pairing~potential \begin{equation}
        \hat{\Delta}_\mathrm{S}(z) = \Delta_0 \hat{\sigma}_0 \Theta(z) \Theta(d-z) ,
    \end{equation}
    with the isotropic \emph{zero-temperature} superconducting energy~gap~$ \Delta_0 $, couples the Bogoliubov--de~Gennes Hamiltonian's electron and hole~blocks. 
    To simplify the analytical description of our system, we take the same Fermi~levels~$ \mu $ and effective carrier~masses~$ m $ in all junction~regions, and approximate the superconducting pairing~potential~$ \hat{\Delta}_\mathrm{S} (z) $ by a Heaviside step~function being nonzero only inside the superconductor and instantly jumping to zero inside the ferromagnets. 
    All other spatial variations of the superconducting~order~parameter, i.e., its more realistic exponential decay in the vicinity of the F/S~boundaries, are fully neglected and would need to be determined from a self-consistent~formulation~\cite{Halterman2001}. 
    However, earlier experimental studies~\cite{Baladie2001,Geers2001} demonstrated that even in F/S/F~junctions in which the superconducting link is much thicker than the Bardeen--Cooper--Schrieffer coherence~length, the current flows still quite uniformly, suggesting that spatial variations of the superconducting order~parameter mostly average out and a self-consistent treatment is not necessary to understand such junctions' generic transport~features. 
    
    Assuming translational invariance parallel to the semiconducting interfaces, the solutions of the Bogoliubov--de~Gennes equation
    \begin{equation}
        \hat{\mathcal{H}}_\mathrm{BdG} \Psi^\sigma (\mathbf{r}) = E \Psi^\sigma (\mathbf{r}) 
    \end{equation}
    factorize into 
    \begin{equation}
        \Psi^\sigma (\mathbf{r}) = \psi^\sigma (z) \mathrm{e}^{\mathrm{i} (\mathbf{k}_\mathbf{\parallel} \cdot \mathbf{r}_\mathbf{\parallel})} ,
    \end{equation}
    where $ \mathbf{k}_\mathbf{\parallel} = [ k_x, \, k_y, \, 0]^\top $~($ \mathbf{r}_\mathbf{\parallel} = [x, \, y, \, 0]^\top $) refers to the in-plane momentum~(position) vector and $ \psi^\sigma (z) $ to the Bogoliubov--de~Gennes equation's most general solution for the effectively one-dimensional scattering~problem along~$ \hat{z} $. 
    Considering an incoming spin-$ \sigma $~electron from the left ferromagnet~[$ \sigma = +(-) 1 $ for spin~up~(spin~down), which effectively indicates a spin parallel~(antiparallel) to~$ \hat{\mathbf{m}}_\mathbf{1} $], the latter is found to read as
    \begin{align}
        \psi^\sigma (z<0) &= \mathrm{e}^{\mathrm{i} k_{z,\mathrm{e}}^\sigma z} \frac{1}{\sqrt{2}} \left[ \sigma \mathrm{e}^{-\mathrm{i} \Phi_1} , \, 1 , \, 0 , \, 0 \right]^\top 
        \nonumber \\
        &\hspace{24 pt} + r_\mathrm{e}^{\sigma,\sigma} \mathrm{e}^{-\mathrm{i} k_{z,\mathrm{e}}^\sigma z} \frac{1}{\sqrt{2}} \left[ \sigma \mathrm{e}^{-\mathrm{i} \Phi_1} , \, 1 , \, 0 , \, 0 \right]^\top 
        \nonumber \\
        &\hspace{24 pt} + r_\mathrm{e}^{\sigma,-\sigma} \mathrm{e}^{-\mathrm{i} k_{z,\mathrm{e}}^{-\sigma} z} \frac{1}{\sqrt{2}} \left[ -\sigma \mathrm{e}^{-\mathrm{i} \Phi_1} , \, 1 , \, 0 , \, 0 \right]^\top 
        \nonumber \\
        &\hspace{24 pt}+ r_\mathrm{h}^{\sigma,-\sigma} \mathrm{e}^{\mathrm{i} k_{z,\mathrm{h}}^{-\sigma} z} \frac{1}{\sqrt{2}} \left[ 0 , \, 0 , \, \sigma \mathrm{e}^{-\mathrm{i} \Phi_1} , \, 1 \right]^\top 
        \nonumber \\
        &\hspace{24 pt} + r_\mathrm{h}^{\sigma,\sigma} \mathrm{e}^{\mathrm{i} k_{z,\mathrm{h}}^\sigma z} \frac{1}{\sqrt{2}} \left[ 0 , \, 0 , \, -\sigma \mathrm{e}^{-\mathrm{i} \Phi_1} , \, 1 \right]^\top 
    \end{align}
    in the left ferromagnet~($ z < 0 $), 
    \begin{align}
        \psi^\sigma (0 < z < d) &= \left( \varepsilon_1^\sigma \mathrm{e}^{\mathrm{i} q_{z,\mathrm{e}} z} + \varepsilon_2^\sigma \mathrm{e}^{-\mathrm{i} q_{z,\mathrm{e}} z} \right) \left[ u , \, 0 , \, v , \, 0 \right]^\top 
        \nonumber \\
        &+ \left( \varepsilon_3^\sigma \mathrm{e}^{\mathrm{i} q_{z,\mathrm{e}} z} + \varepsilon_4^\sigma \mathrm{e}^{-\mathrm{i} q_{z,\mathrm{e}} z} \right) \left[ 0 , \, u , \, 0 , \, v \right]^\top 
        \nonumber \\
        &+ \left( \eta_1^\sigma \mathrm{e}^{-\mathrm{i} q_{z,\mathrm{h}} z} + \eta_2^\sigma \mathrm{e}^{\mathrm{i} q_{z,\mathrm{h}} z} \right) \left[ v , \, 0 , \, u , \, 0 \right]^\top 
        \nonumber \\
        &+ \left( \eta_3^\sigma \mathrm{e}^{-\mathrm{i} q_{z,\mathrm{h}} z} + \eta_4^\sigma \mathrm{e}^{\mathrm{i} q_{z,\mathrm{h}} z} \right) \left[ 0 , \, v , \, 0 , \, u \right]^\top 
    \end{align}
    in the superconducting link~($ 0 < z < d $), and accordingly 
    \begin{align}
        \psi^\sigma (z > d) &= t_\mathrm{e}^{\sigma,\sigma} \mathrm{e}^{\mathrm{i} k_{z,\mathrm{e}}^\sigma z} \frac{1}{\sqrt{2}} \left[ \sigma \mathrm{e}^{-\mathrm{i} \Phi_2} , \, 1 , \, 0 , \, 0 \right]^\top 
        \nonumber \\
        &\hspace{18 pt} + t_\mathrm{e}^{\sigma,-\sigma} \mathrm{e}^{\mathrm{i} k_{z,\mathrm{e}}^{-\sigma} z} \frac{1}{\sqrt{2}} \left[ -\sigma \mathrm{e}^{-\mathrm{i} \Phi_2} , \, 1 , \, 0 , \, 0 \right]^\top 
        \nonumber \\ 
        &\hspace{18 pt} + t_\mathrm{h}^{\sigma,-\sigma} \mathrm{e}^{-\mathrm{i} k_{z,\mathrm{h}}^{-\sigma} z} \frac{1}{\sqrt{2}} \left[ 0 , \, 0 , \, \sigma \mathrm{e}^{-\mathrm{i} \Phi_2} , \, 1 \right]^\top 
        \nonumber \\
        &\hspace{18 pt} + t_\mathrm{h}^{\sigma,\sigma} \mathrm{e}^{-\mathrm{i} k_{z,\mathrm{h}}^{-\sigma} z} \frac{1}{\sqrt{2}} \left[ 0 , \, 0 , \, -\sigma \mathrm{e}^{-\mathrm{i} \Phi_2} , \, 1 \right]^\top 
    \end{align}
    in the right ferromagnet~($ z > d $). 
    The $ \hat{z} $-projected wave~vectors of spin-$ \sigma $ electrons and holes in the ferromagnets are given by 
    \begin{align}
        k_{z,\mathrm{e}}^\sigma &= \sqrt{k_\mathrm{F}^2 + 2m/\hbar^2 (E + \sigma \Delta_\mathrm{XC} / 2 ) - |\mathbf{k}_\mathbf{\parallel}|^2} 
        \intertext{and} 
        k_{z,\mathrm{h}}^\sigma &= \sqrt{k_\mathrm{F}^2 + 2m/\hbar^2 (-E + \sigma \Delta_\mathrm{XC} / 2 ) - |\mathbf{k}_\mathbf{\parallel}|^2} ,
    \end{align}
    respectively, whereas we obtain 
    \begin{align}
        q_{z,\mathrm{e}} &= \sqrt{k_\mathrm{F}^2 + 2m/\hbar^2 \sqrt{E^2 - \Delta_0^2} - | \mathbf{k}_\mathbf{\parallel} |^2}
        \intertext{and}
        q_{z,\mathrm{h}} &= \sqrt{k_\mathrm{F}^2 - 2m/\hbar^2 \sqrt{E^2 - \Delta_0^2} - | \mathbf{k}_\mathbf{\parallel} |^2} 
    \end{align}
    for electronlike and holelike quasiparticles inside the superconducting link; $ k_\mathrm{F} = \sqrt{2m\mu} / \hbar $ denotes the Fermi~wave~vector. 
    Finally, the Bardeen--Cooper--Schrieffer coherence~factors can be written as
    \begin{align}
        u &= \sqrt{ \frac{1}{2} \left( 1 + \frac{\sqrt{E^2 - \Delta_0^2}} {E} \right) } \, ,
        \intertext{as well as}
        v &= \sqrt{ 1 - u^2 } .
    \end{align}

    The given states account for all scattering~processes that incident electrons may undergo at the semiconductor~interfaces, including also the possibility of spin-flip scattering caused by the interfacial SOCs. 
    Apart from spin-conserving and spin-flip specular~(normal)~reflections~(with amplitudes $ r_\mathrm{e}^{\sigma,\sigma} $ and $ r_\mathrm{e}^{\sigma,-\sigma} $), we need to pay special attention to spin-conserving~(``conventional'') and spin-flip~(``unconventional'') Andreev~reflections~(with amplitudes~$ r_\mathrm{h}^{\sigma,-\sigma} $ and $ r_\mathrm{h}^{\sigma,\sigma} $), which usually induce superconducting~order in the ferromagnetic~electrodes through proximity and lead thereby to numerous unique physical characteristics in superconducting magnetic junctions. 
    Regarding transmissions~(tunnelings) into the right ferromagnet, we need to distinguish between electron transmissions~(with amplitudes~$ t_\mathrm{e}^{\sigma,\sigma} $ and $ t_\mathrm{e}^{\sigma,-\sigma} $) on the one hand and hole transmissions~(with amplitudes~$ t_\mathrm{h}^{\sigma,-\sigma} $ and $ t_\mathrm{h}^{\sigma,\sigma} $) on the other.

    To determine the unknown scattering~amplitudes, we apply the interfacial boundary~conditions~(at~$ z=0 $ and $ z=d $) 
    \footnotesize
    \begin{equation}
        \psi^\sigma (z=0_-) = \psi^\sigma (z=0_+) , \quad \quad \psi^\sigma (z=d_-) = \psi^\sigma (z=d_+) ,
    \end{equation}
    \begin{multline}
        \frac{\hbar^2}{2m} \left[ \frac{\mathrm{d}}{\mathrm{d}z} \hat{\eta} \psi^\sigma (z=0_+) - \frac{\mathrm{d}}{\mathrm{d}z} \hat{\eta} \psi^\sigma (z=0_-) \right]  \\
        = \left[ \begin{matrix} \boldsymbol{\Omega}_\mathbf{L} \cdot \hat{\boldsymbol{\sigma}} + V_\mathrm{L} d_\mathrm{L} \hat{\sigma}_0 & \hat{0} \\ \hat{0} & -( \boldsymbol{\Omega}_\mathbf{L} \cdot \hat{\boldsymbol{\sigma}} + V_\mathrm{L} d_\mathrm{L} \hat{\sigma}_0 ) \end{matrix} \right] \psi^\sigma (z=0_+) ,
    \end{multline}
    \normalsize
    as well as
    \footnotesize
    \begin{multline}
        \frac{\hbar^2}{2m} \left[ \frac{\mathrm{d}}{\mathrm{d}z} \hat{\eta} \psi^\sigma (z=d_+) - \frac{\mathrm{d}}{\mathrm{d}z} \hat{\eta} \psi^\sigma (z=d_-) \right]  \\
        = \left[ \begin{matrix} \boldsymbol{\Omega}_\mathbf{R} \cdot \hat{\boldsymbol{\sigma}} + V_\mathrm{R} d_\mathrm{R} \hat{\sigma}_0 & \hat{0} \\ \hat{0} & -( \boldsymbol{\Omega}_\mathbf{R} \cdot \hat{\boldsymbol{\sigma}} + V_\mathrm{R} d_\mathrm{R} \hat{\sigma}_0 ) \end{matrix} \right] \psi^\sigma (z=d_+) 
    \end{multline}
    \normalsize
    to the scattering~states and numerically solve the resulting linear system of equations~(at a given spin~$ \sigma $; $ \hat{\eta} = \mathrm{diag} [1, \, 1, \, -1, \, -1] $ and $ \hat{0} $ abbreviates the $ 2 \times 2 $~zero~matrix).

    Assuring charge~conservation~\footnote{Note that one needs to be careful when evaluating electrical currents from the Blonder--Tinkham--Klapwijk~approach in junctions with more than two different regions to ensure that the calculated current is indeed conserved; see the thorough discussion in~Ref.~\cite{Dong2003}. }, and taking both the electron and hole contributions into account~\cite{Lambert1991,Bozovic2002,*Bozovic2005,Yamashita2003,Dong2003}, the tunneling~conductance at \emph{zero temperature} and bias~voltage~$ V $ can be evaluated from 
    \begin{align}
        G &= \frac{\mathcal{A} e^2 k_\mathrm{F}^2}{2 \pi^2 h} \sum_{\sigma = \mp 1} \int \mathrm{d}^2 \mathbf{k}_\mathbf{\parallel} \, \left[ \left| t_\mathrm{e}^{\sigma,\sigma} \right|^2  \right. \nonumber \\
        &\hspace{75 pt} + \mathrm{Re} \left( \frac{k_{z,\mathrm{e}}^{-\sigma}}{k_{z,\mathrm{e}}^\sigma} \right) \left| t_\mathrm{e}^{\sigma,-\sigma} \right|^2 \nonumber \\
        &\hspace{75 pt} + \mathrm{Re} \left( \frac{k_{z,\mathrm{h}}^{-\sigma}}{k_{z,\mathrm{e}}^\sigma} \right) \left| r_\mathrm{h}^{\sigma,-\sigma} \right|^2 \nonumber \\
        &\hspace{75 pt} + \left. \left. \mathrm{Re} \left( \frac{k_{z,\mathrm{h}}^{\sigma}}{k_{z,\mathrm{e}}^\sigma} \right) \left| r_\mathrm{h}^{\sigma,\sigma} \right|^2 \right] \right|_{E = eV} ,
        \label{Eq_Conductance}
    \end{align}
    where $ \mathcal{A} $ indicates the contact cross-section~area, $ e $ is the positive elementary~charge, $ h = 2\pi \hbar $ corresponds to Planck's~constant, and taking the real~parts~($ \mathrm{Re} \ldots $) of the wave-vector~ratios ensures that only the contributions originating from \emph{propagating} states are included in the conductance~calculation. 
    Interestingly, the tunneling~conductance of superconducting F/S/F~junctions is therefore not only governed by the usual tunneling~electrons~(electron~transmissions)---as is the case in normal-state F/N/F~junctions according to Julli\`{e}re's model~\cite{Julliere1975}---but moreover impacted by the crucial Andreev-reflection~process, which additionally transfers electrons in terms of supercurrent-carrying Cooper~pairs across the superconducting junction~link and will most likely give rise to unforeseen physical phenomena.

    \section{General conductance~features and signatures of induced triplet pairing  \label{Sec_General}}

    To analyze the most fundamental features and tunability of the tunneling~conductance, we numerically evaluate Eq.~\eqref{Eq_Conductance} for realistic junction~parameters. 
    More specifically, we assume that both semiconducting tunneling~barriers introduce the same weak potential~scattering described by the dimensionless Blonder--Tinkham--Klapwijk parameters~\cite{Blonder1982} $ Z_\mathrm{L} = (2m V_\mathrm{L} d_\mathrm{L}) / (\hbar^2 k_\mathrm{F}) = 1 = (2m V_\mathrm{R} d_\mathrm{R}) / (\hbar^2 k_\mathrm{F}) = Z_\mathrm{R} $, which would correspond, for~example, to barrier~heights~$ V_\mathrm{L} = V_\mathrm{R} \approx 0.75 \, \mathrm{eV} $ and widths~$ d_\mathrm{L} = d_\mathrm{R} = 0.40 \, \mathrm{nm} $~(substituting the typical Fermi~wave~vector~$ k_\mathrm{F} \approx 8 \times 10^7 \, \mathrm{cm}^{-1} $ of iron~\cite{Martinez2020} and approximating the effective carrier~mass by the free-electron~mass). 
    These barriers mimic reduced interfacial transparencies that could stem, e.g., from scattering due to imperfect interfaces or strongly differing Fermi~levels~(electron~densities) in the ferromagnetic and superconducting junction~regions~\cite{Zutic2000}. 
    The strengths of the interfacial Rashba and Dresselhaus~SOCs are quantified by the dimensionless measures~$ \lambda^\alpha_\mathrm{L} = (2m\alpha_\mathrm{L}) / \hbar^2 $, $ \lambda^\alpha_\mathrm{R} = (2m\alpha_\mathrm{R}) / \hbar^2 $, $ \lambda^\beta_\mathrm{L} = (2m\beta_\mathrm{L}) / \hbar^2 $, and $ \lambda^\beta_\mathrm{R} = (2m\beta_\mathrm{R}) / \hbar^2 $. 
    For simplicity, we assume that both semiconducting~interfaces are identical, i.e., they are characterized by the same Rashba and Dresselhaus SOC~parameters, respectively. 
    Rashba~(Dresselhaus)~parameters of~$ \lambda^\alpha_\mathrm{L} = \lambda^\alpha_\mathrm{R} = 0.5 $~($ \lambda^\beta_\mathrm{L} = \lambda^\beta_\mathrm{R} = 0.5 $) correspond then to bare Rashba~(Dresselhaus)~SOCs of~$ \alpha_\mathrm{L} = \alpha_\mathrm{R} \approx 1.9 \, \mathrm{eV} \, \text{\AA}^2 $~($ \beta_\mathrm{L} = \beta_\mathrm{R} \approx 1.9 \, \mathrm{eV} \, \text{\AA}^2 $, connecting $ \beta_\mathrm{L} \approx Z_\mathrm{L} k_\mathrm{F} \gamma_\mathrm{L} $ and $ \beta_\mathrm{R} \approx Z_\mathrm{R} k_\mathrm{F} \gamma_\mathrm{R} $ with the barriers' cubic Dresselhaus~parameters~$ \gamma_\mathrm{L} $ and $ \gamma_\mathrm{R} $ that are typically given in the literature~\cite{Fabian2007,MatosAbiague2009}). 
    Recall that a $ 1.7 \, \mathrm{nm} $~thick MgO~barrier was found to induce Rashba~SOCs up to~$ \alpha \approx 4.6 \, \mathrm{eV} \, \text{\AA}^2 $ in Fe/MgO/V~junctions~\cite{Martinez2020}, while AlP~barriers of the given heights and widths would indeed raise Dresselhaus~SOCs of~$ \beta \approx 1.7 \, \mathrm{eV} \, \text{\AA}^2 $. 
    Finally, the spin~polarization of the (identical) ferromagnets is determined by the dimensionless variable~$ P = (\Delta_\mathrm{XC} / 2) / \mu $. 
    The Fermi~level~$ \mu $ is typically much larger than the superconducting~gap~$ \Delta_0 $ and the excitation~energies~$ E $, motivating $ \mu = 10^3 \Delta_0 $ as a reasonable assumption for our calculations. 
    All tunneling-conductance~values discussed throughout this paper are normalized to the respective normal-state~conductance~$ G_\mathrm{N} = (\mathcal{A} e^2 k_\mathrm{F}^2 ) / (2\pi^2 \hbar) $ of an ideal~(perfectly transparent) N/N/N~junction.

    \begin{figure}
        \centering
        \includegraphics[width=0.45\textwidth]{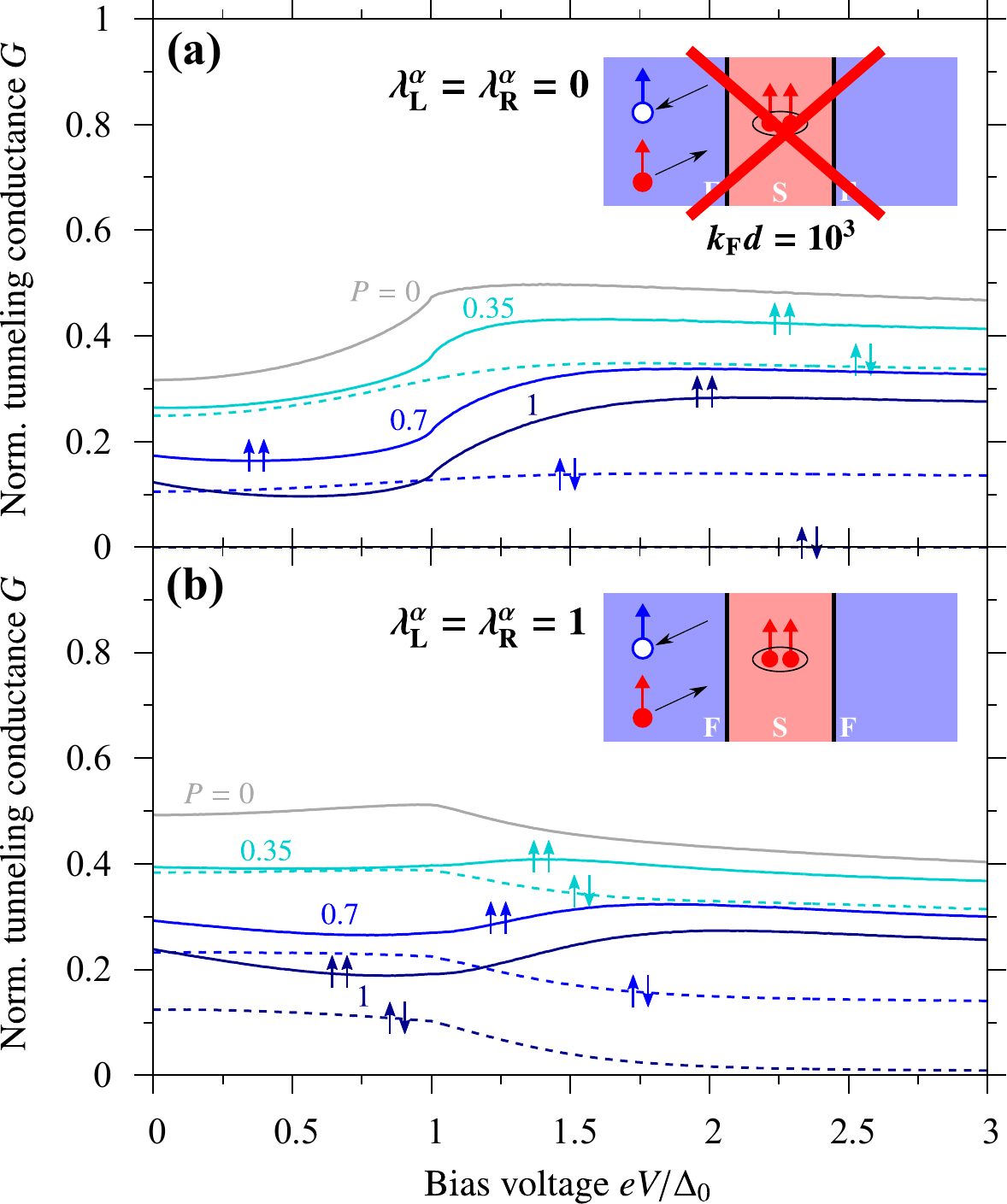}
        \caption{Calculated tunneling~conductance~$ G $ as a function of the applied bias~voltage~$ V $ and for different indicated spin~polarizations~$ P $ of the ferromagnetic electrodes, considering a \emph{``thin'' superconducting link} of thickness~$ d = 10^3 / k_\mathrm{F} $. 
            \emph{Solid lines} correspond to \emph{parallel magnetization orientations}~(both ferromagnets are magnetized along the~$ \hat{y} $-direction; $ \uparrow \uparrow $), whereas \emph{dashed lines} indicate \emph{antiparallel magnetization orientations}~(the left ferromagnet is magnetized along~$ \hat{y} $ and the right along~$ -\hat{y} $; $ \uparrow \downarrow $). 
            (a)~In the absence of interfacial SOCs~($ \lambda^\alpha_\mathrm{L} = \lambda^\alpha_\mathrm{R} = \lambda^\beta_\mathrm{L} = \lambda^\beta_\mathrm{R} = 0 $), unconventional Andreev~reflections are forbidden~(see illustration), and the conductance is fully determined by conventional Andreev~reflections and electron transmissions. 
            (b)~Unconventional Andreev-reflection~contributions~(see illustration) at moderate interfacial Rashba~SOCs~($ \lambda^\alpha_\mathrm{L} = \lambda^\alpha_\mathrm{R} = 1 $) may significantly enhance the subgap tunneling~conductance~(at~$ eV < \Delta_0 $). 
        }
        \label{Fig_GeneralConductanceWithSOC_Short}
    \end{figure}

    Figure~\ref{Fig_GeneralConductanceWithSOC_Short} illustrates the tunneling~conductance of a F/S/F~junction comprising a ``thin'' superconducting link of thickness~$ d = 10^3 / k_\mathrm{F} $~($ \approx 125 \, \mathrm{nm} $ if~$ k_\mathrm{F} \approx 8 \times 10^7 \, \mathrm{cm}^{-1} $) once in the absence of interfacial SOCs and once if the moderate Rashba~SOCs~$ \lambda^\alpha_\mathrm{L} = \lambda^\alpha_\mathrm{R} = 1 $ are present. 
    In both cases, we consecutively increase the ferromagnets' spin~polarization from~$ P = 0 $~(N/S/N~junction; gray color) to~$ P = 1 $~(half-metallic~F/S/half-metallic~F; dark-blue color), and distinguish parallel~(both magnetizations point along~$ \hat{y} $) from antiparallel~(the left ferromagnet's magnetization points along~$ \hat{y} $ and the right one's along~$ -\hat{y} $) magnetizations.

    Let us first focus on the situation without interfacial SOCs~[see~Fig.~\ref{Fig_GeneralConductanceWithSOC_Short}(a)], in which spin-flip~scatterings---i.e., also the crucial unconventional Andreev~reflections---are completely forbidden, and all electrical transport is governed by electron~transmissions~(hole~transmissions are negligible) and conventional Andreev~reflections. 
    However, as the superconducting~link is rather thin when compared to usual superconducting coherence~lengths~(in the micron range~\cite{Kittel1996}), also conventional Andreev~reflections are heavily suppressed---particularly at large spin~polarizations, at which the through conventional Andreev~reflections proximity-induced superconducting~order inside the ferromagnets becomes negligibly tiny anyway due to these metals' small coherence~lengths~\cite{Yang2005}. 
    As a consequence, thin superconducting~links act mostly like rectangular potential~barriers of height~$ \Delta_0 $ and width~$ d $. 
    Since $ \Delta_0 \ll \mu $, even subgap-energy~electrons~($ eV < \Delta_0 $) can tunnel across the superconducting link with considerably large probabilities, entailing nonzero subgap tunneling~conductances. 
    Nevertheless, electron~transmissions happen of~course much more likely at energies above the barrier~height~(at~$ eV \geq \Delta_0 $), explaining the enhanced tunneling~conductances there. 
    Increasing the ferromagnets' spin~polarization monotonically decreases the tunneling-conductance~amplitudes, as less minority-spin~electrons are then involved in tunneling and can contribute to transport. 
    According to Julli\`{e}re's~model~\cite{Julliere1975}, switching the \emph{relative} magnetization~orientations from the parallel to their antiparallel configuration~(at~$ P \neq 0 $) notably damps the probability for electron~transmissions and thereby the tunneling~conductance---most extreme in the half-metallic case in which no electrons can tunnel between oppositely magnetized ferromagnets and huge MR~ratios are expected.

    Second, we explore the effects associated with interfacial Rashba~SOCs~[see~Fig.~\ref{Fig_GeneralConductanceWithSOC_Short}(b)]. 
    Although conventional Andreev~reflections are still strongly suppressed, the tunneling~conductances in the subgap bias-voltage~regime are remarkably enhanced by the present SOCs. 
    This enhancement stems from the now additionally possible unconventional Andreev~reflections~\footnote{The tunneling-conductance~enhancement in the case of N/S/N~junctions~(i.e., at~$ P=0 $) cannot be attributed to unconventional Andreev~reflections, which become most pronounced in \emph{ferromagnetic} junctions. Instead, the Rashba~SOC terms in the single-particle~Hamiltonian~[recall~Eq.~(\ref{Eq_SingleHamiltonian})] always partially compensate the potential~barrier~(for appropriate~$ |\mathbf{k}_\parallel| $ and~$ \sigma $), and lead thus to more electron~transmissions and thereby a larger tunneling~conductance. This can cause a conductance~enhancement even in N/S/N~junctions although unconventional Andreev~reflections are strongly suppressed there.}, which are known to induce sizable superconducting triplet~pairings even in strongly spin-polarized ferromagnets~(and even though the superconducting link in our case is quite thin). 
    The unconventional Andreev-reflection contributions become maximal in magnitude close to the chosen Rashba~SOC strengths~$ \lambda^\alpha_\mathrm{L} = \lambda^\alpha_\mathrm{R} = 1 $~(but are still slightly smaller than the electron-transmission~parts), can cause finite (subgap) tunneling~conductances even in the case of half-metallic ferromagnetic electrodes, and are only merely affected by flipping the ferromagnets' \emph{relative} magnetization~orientations---giving reasoning for the smaller~(compared to the case without SOCs) absolute conductance~changes when switching between parallel and antiparallel configurations~(see our discussions in~Sec.~\ref{Sec_TMR}). 
    Increasing the bias~voltage to values above the superconducting gap~($ eV \geq \Delta_0 $), unconventional Andreev~reflections become more unlikely and we essentially recover the purely by electron~transmissions dominated transport~regime at~$ eV \gg \Delta_0 $. 
    However, we wish to stress that the interfacial SOCs furthermore act like additional deltalike potential~barriers to electron~transmissions~[recall the Hamiltonian in~Eq.~\eqref{Eq_SingleHamiltonian}], eventually leading to even smaller (normal-state)  tunneling~conductances than in the absence of SOCs.

    \begin{figure}
        \centering
        \includegraphics[width=0.45\textwidth]{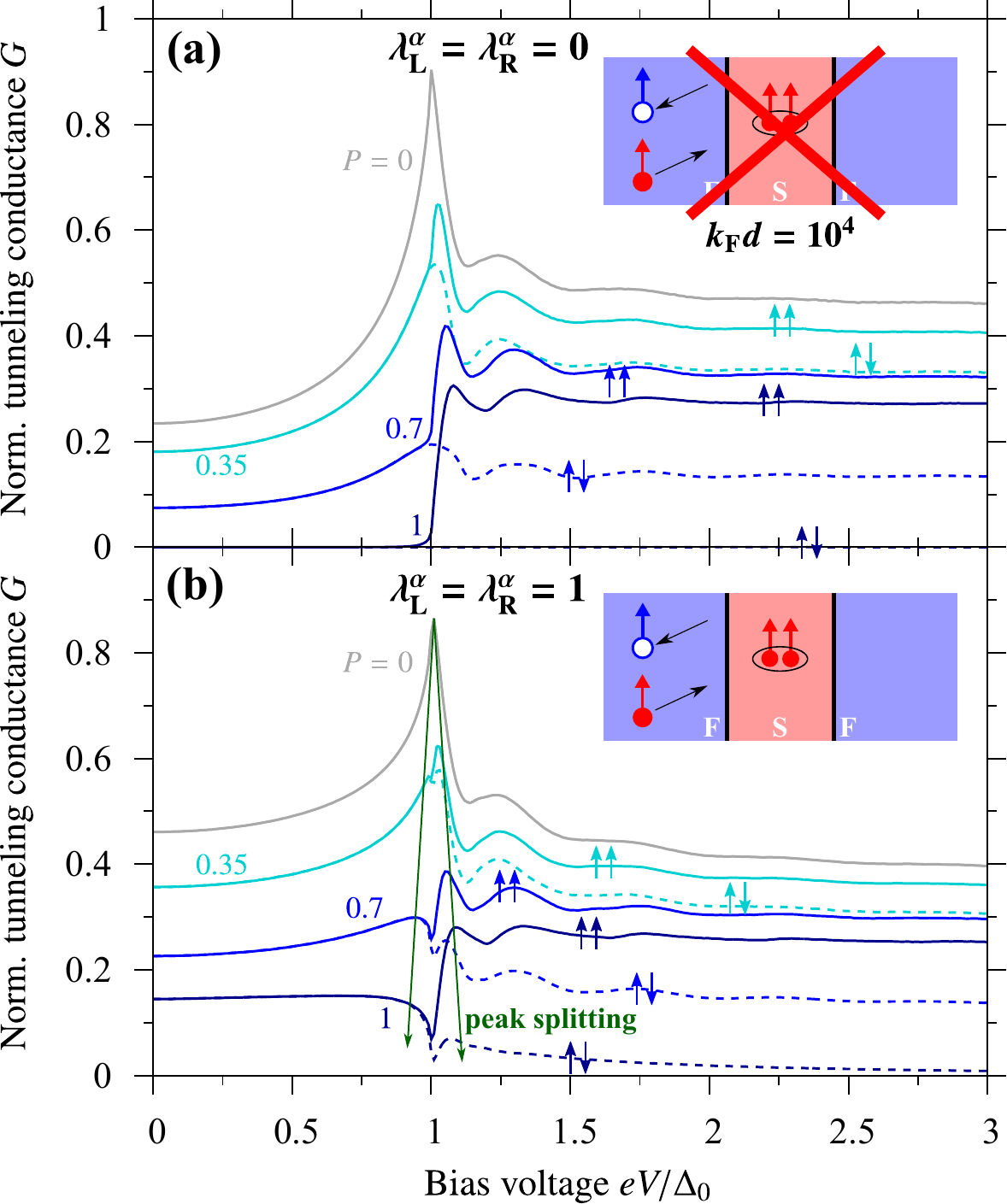}
        \caption{Same calculations as in~Fig.~\ref{Fig_GeneralConductanceWithSOC_Short}, but assuming a \emph{``thick'' superconducting link} of thickness~$ d = 10^4 / k_\mathrm{F} $. 
            The conductance~spectra in the presence of interfacial Rashba~SOCs are now \emph{dominated} by unconventional Andreev~reflections, effectively inducing nonzero superconducting triplet~gaps close to the superconducting interfaces~(due to the formation of polarized spin-triplet Cooper~pairs) that are evident in terms of split gap-edge conductance double~peaks around $ eV = \Delta_0 $. 
        }
        \label{Fig_GeneralConductanceWithSOC_Long}
    \end{figure}

    Next, we investigate the tunneling~conductance of a F/S/F~junction that contains a ``thick'' superconducting link of thickness~$ d = 10^4 / k_\mathrm{F} $~($ \approx 1.25 \, \upmu \mathrm{m} $ if~$ k_\mathrm{F} \approx 8 \times 10^7 \, \mathrm{cm}^{-1} $); see~Fig.~\ref{Fig_GeneralConductanceWithSOC_Long}. 
    Such junctions are probably of much greater relevance to future experimental studies since their thicker superconducting regions inherently entail much larger Andreev-reflection~probabilities and their conductance~spectra simultaneously reveal the superconductor's most fundamental spectroscopic~fingerprints~(gap).

    We start again analyzing the case without interfacial SOCs~[see~Fig.~\ref{Fig_GeneralConductanceWithSOC_Long}(a)]. 
    In~fact, electron~transmissions are now completely forbidden at~$ eV < \Delta_0 $ and the subgap tunneling~conductance is fully describable through the properties of conventional Andreev~reflections~(implicitly also carrying all necessary information about specular~reflections if nonzero potential~barriers are present). 
    As we mentioned before, Andreev~reflections are only merely affected by switching the ferromagnets' \emph{relative} magnetization~orientations. 
    Therefore, the subgap conductances for parallel and antiparallel magnetizations are (nearly) equal in magnitude as long as the tunneling~conductance is fully dominated by Andreev~reflections.

    At the superconducting gap~edge~($ eV = \Delta_0 $), the tunneling~conductance of a nonmagnetic N/S/N~junction always reflects a sharp conductance~peak, which indicates the superconductor's density-of-states coherence~peak and from which the superconducting energy~gap can be estimated through transport~experiments. 
    With increasing spin~polarization in a ferromagnetic junction, the conductance~peak flattens and its position moves to energies slightly above the gap---though one can still quite reliably estimate the value of the superconducting gap from the peak~position. 
    Above the gap~($ eV > \Delta_0 $), the tunneling~conductance reveals unique oscillations, which are damped out with increasing voltage and finally disappear when approaching the normal-state transport~regime at~$ eV \gg \Delta_0 $. 
    These oscillations reflect the coherency of electrical transport through F/S/F~junctions' superconducting links.
    Coherent interference of incoming and outgoing quasiparticles~(that underwent multiple reflections at the S/F~interfaces) basically leads to Andreev-reflection and electron-transmission probabilities that strongly oscillate as functions of the excitation~energy~$ E=eV $ and the link~thickness~$ d $; for the latter reason, the oscillations are usually referred to as \emph{geometrical oscillations}.

    An earlier work~\cite{Bozovic2002,*Bozovic2005} concluded that Andreev~reflections are strongly suppressed and coherent electron~transmissions become concurrently most likely whenever 
    \begin{equation}
        (q_{z,\mathrm{e}} - q_{z,\mathrm{h}}) d = 2\pi n ,
    \end{equation}
    where $ n $ is an integer and, for~simplicity, the effects of tunneling~barriers~(and large spin polarizations) were neglected. 
    Inspecting our numerical results shows indeed that finite-height tunneling~barriers and (large) spin~polarizations only barely impact the conductance~oscillations. 
    Similar oscillations would actually be expected to likewise occur in junctions with thinner superconducting links, but the oscillation~period is so large there~(owing to the much smaller~$ d $) that we did not resolve them within the bias-voltage~range chosen in~Fig.~\ref{Fig_GeneralConductanceWithSOC_Short}. 
    Regarding the tunneling-conductance~amplitudes, increasing spin~polarization decreases both the Andreev-reflection and electron-transmission contributions, and thereby suppresses the conductance. In the half-metallic~case, the subgap~conductance vanishes now even in the parallel magnetization~configuration since all subgap transport is governed by conventional Andreev~reflections, which are no longer possible if only majority-spin~electrons are available.

    Finite Rashba~SOCs at the semiconducting interfaces additionally allow for the crucial unconventional Andreev~reflections. 
    While the general conductance~features far below and far above the superconducting~gap~$ \Delta_0 $ are the same as we thoroughly discussed earlier---including the conductance~increase in the subgap~region, the conductance~decrease at~$ eV \gg \Delta_0 $, finite tunneling~conductances in the half-metallic limit, and the geometrical oscillations at~$ eV > \Delta_0 $---the most puzzling feature arises in the vicinity of the gap~edge itself. 
    Increasing the ferromagnets' spin~polarization splits the conductance~peak that we attributed to the gap-edge density-of-states coherence~peak into two distinct peaks---one located below and the other above the gap~energy~$ \Delta_0 $. 
    This peak~splitting becomes most pronounced as the spin~polarization approaches the half-metallic~limit~($ P \to 1 $) and the unconventional Andreev-reflection conductance~contribution becomes considerably large when compared to conventional Andreev~reflections and electron~transmissions. 
    The latter observation might serve as a hint that the conductance-peak~splitting and the peculiar unconventional Andreev-reflection~process must be intimately connected.

    \begin{figure}
        \centering
        \includegraphics[width=0.45\textwidth]{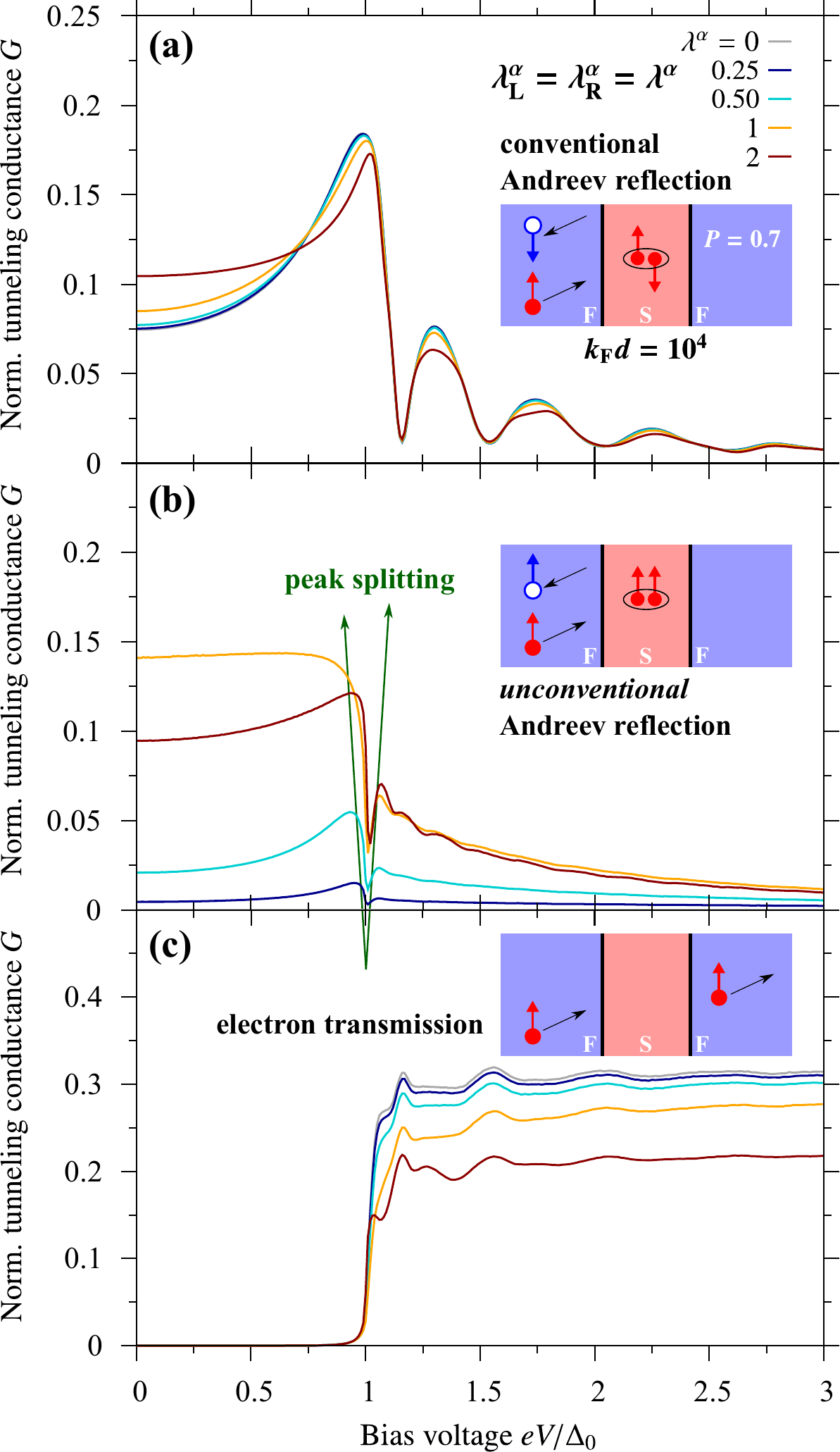}
        \caption{Calculated tunneling~conductance~$ G $ as a function of the applied bias~voltage~$ V $ and for different indicated interfacial Rashba~SOC~strengths~$ \lambda^\alpha_\mathrm{L} = \lambda^\alpha_\mathrm{R} = \lambda^\alpha $~(Dresselhaus SOC is not present; $ \lambda^\beta_\mathrm{L} = \lambda^\beta_\mathrm{R} = 0 $), considering iron as ferromagnetic~electrodes~(spin~polarization~$ P = 0.7 $), both ferromagnets magnetized along~$ \mp \hat{y} $, and a \emph{``thick'' superconducting link} of thickness~$ d = 10^4 / k_\mathrm{F} $. 
            The individually presented conductance contributions stem from (a)~conventional Andreev~reflections, (b)~unconventional Andreev~reflections, and (c)~electron~transmissions. 
        }
        \label{Fig_GeneralConductanceWithSOC_Long_Resolved}
    \end{figure}

    To resolve this connection, Fig.~\ref{Fig_GeneralConductanceWithSOC_Long_Resolved} illustrates the individual conductance~contributions originating from conventional Andreev~reflections, unconventional Andreev~reflections, and electron~transmissions. 
    For brevity, we just discuss the case of parallel magnetizations~(antiparallel magnetizations cause similar physics, but of~course different conductance~amplitudes) and focus on the representative spin~polarization~$ P=0.7 $ of iron electrodes, varying now instead the Rashba~SOC~strengths. 
    Note that the conventional Andreev-reflection and electron-transmission~parts clearly reflect the aforementioned geometrical oscillations with mutually suppressed~(enhanced) Andreev-reflection~(electron-transmission) probabilities at $ eV > \Delta_0 $. 
    The physics becomes nevertheless most interesting close to the gap~edge. 
    While conventional Andreev~reflections still cause dominant conductance~maxima~(``peaks'') at~$ \Delta_0 $---slightly sharpened and shifted with increasing Rashba~SOCs---it is indeed the unconventional Andreev-reflection~process that manifests itself in terms of split conductance~double~peaks located at energies slightly below and above~$ \Delta_0 $. 
    This peak~splitting becomes again most clearly visible when unconventional Andreev~reflection dominates subgap transport~(i.e., close to $ \lambda^\alpha_\mathrm{L} = \lambda^\alpha_\mathrm{R} = 1 $; recall that the same happened at large spin~polarizations). 
    Similarly to conventional Andreev~reflections, which microscopically induce superconducting \emph{singlet} correlations into the ferromagnets close to the interfaces, unconventional Andreev~reflections introduce spin-polarized \emph{triplet} correlations. 
    As a consequence, the gap-edge density-of-states coherence peak splits into two peaks corresponding to $ \Delta_\mathrm{s} - \Delta_\mathrm{t} $ and $ \Delta_\mathrm{s} + \Delta_\mathrm{t} $, accordingly; $ \Delta_\mathrm{s} \approx \Delta_0 $~($ \Delta_\mathrm{t} $) indicates the superconducting gap due to singlet~(triplet) pairing; note, however, that the triplet gap is small when compared to its singlet counterpart, as triplet~pairing is in our case only induced through the interfacial SOCs~(i.e., $ \Delta_\mathrm{t} \ll \Delta_\mathrm{s} $). 
    Measuring the junction's tunneling~conductance probes therefore directly the competing mixture of singlet and triplet correlations at the same time, and detects gap-edge conductance double~peaks as an indirect signature of superconducting triplet pairings, which might help to identify the dominant pairing~mechanism in upcoming transport~studies.

    \section{Magnetic tunability of conductance~features  \label{Sec_MAAR}}

    Apart from the enhancement of the subgap~conductance and the coherence-peak~(conductance-peak) splitting, interfacial SOC gives typically rise to unique transport~magnetoanisotropies, i.e., rotating the magnetization~direction of (at~least) one ferromagnet considerably alters the conductance~amplitudes in the presence of interfacial SOCs. 
    While out-of-plane magnetization~rotations~(in a plane perpendicular to the semiconducting barriers) already cause magnetoanisotropic conductances if the barriers induce either Rashba or Dresselhaus SOCs, in-plane magnetoanisotropies require interfering Rashba \emph{and} Dresselhaus~SOCs. 
    As out-of-plane magnetization~directions are not realistic in spin-valve MR~geometries, we focus on the in-plane case.

    \begin{figure}
        \centering
        \includegraphics[width=0.45\textwidth]{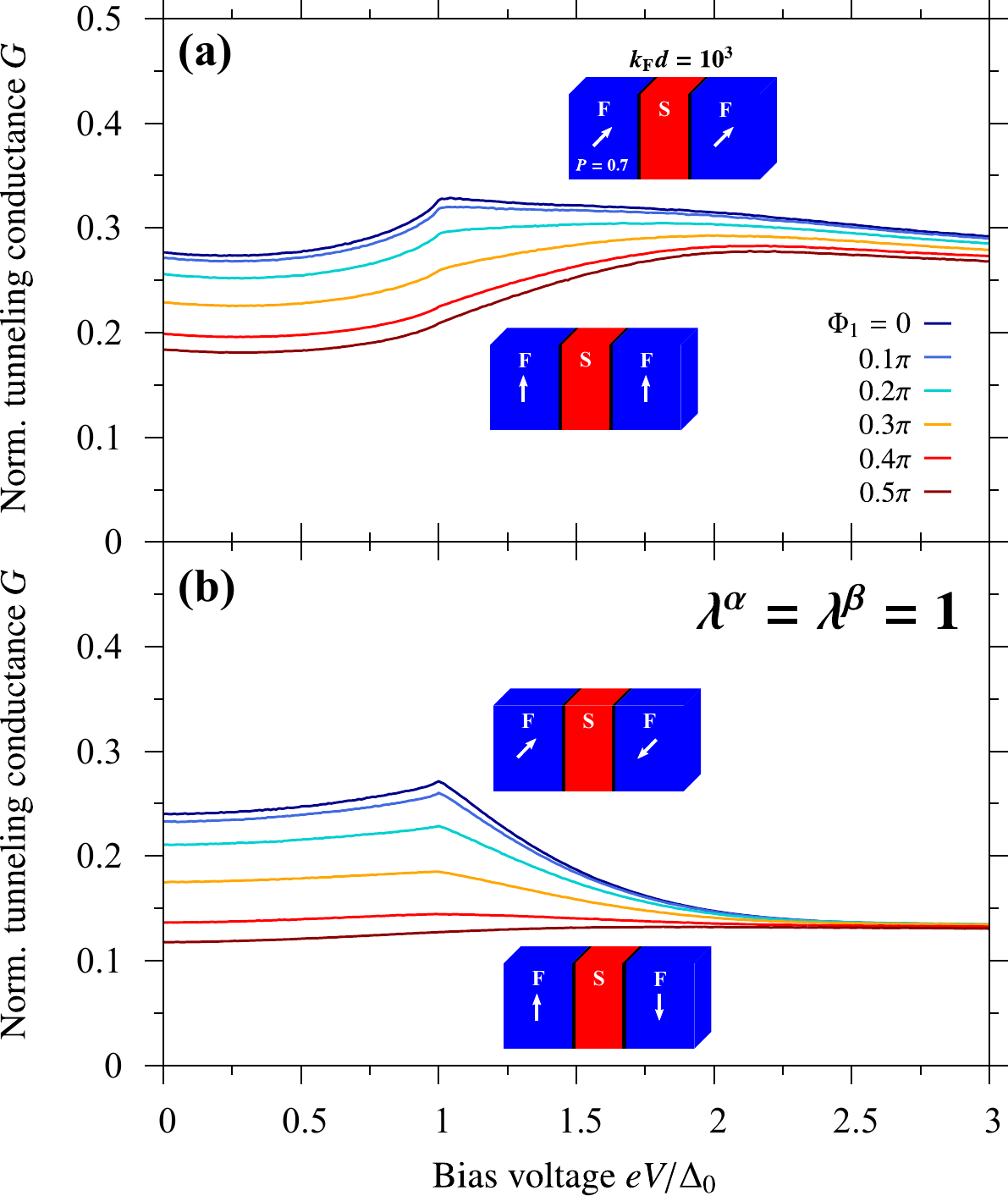}
        \caption{Calculated tunneling~conductance~$ G $ as a function of the applied bias~voltage~$ V $ and for different indicated magnetization~orientations covering the (a)~parallel and (b)~antiparallel magnetized configurations; $ \Phi_1 = 0 $~$ (0.5 \pi) $ corresponds to a magnetization in the left ferromagnet that points along $ \hat{x} $~($ \hat{y} $), as illustrated. 
            The spin~polarization of the ferromagnets is~$ P = 0.7 $, the interfacial SOC~strengths are~$ \lambda^\alpha_\mathrm{L} = \lambda^\alpha_\mathrm{R} = \lambda^\beta_\mathrm{L} = \lambda^\beta_\mathrm{R} = 1 $, and the \emph{``thin'' superconducting link} has the thickness~$ d = 10^3 / k_\mathrm{F} $. 
        }
        \label{Fig_GeneralConductanceWithSOC_Short_MagnetizationTriplet}
    \end{figure}

    Figure~\ref{Fig_GeneralConductanceWithSOC_Short_MagnetizationTriplet} shows the tunneling~conductance of a F/S/F~junction with a ``thin'' superconducting link~($ d = 10^3 / k_\mathrm{F} $) considering the moderate (and equal in magnitude) Rashba and Dresselhaus~SOCs~$ \lambda^\alpha_\mathrm{L} = \lambda^\alpha_\mathrm{R} = \lambda^\beta_\mathrm{L} = \lambda^\beta_\mathrm{R} = 1 $ and rotating the magnetization~(once in the parallel and once in the antiparallel configuration) from the $ \hat{x} $- to the $ \hat{y} $-direction~(from~$ \Phi_1 = 0 $ to $ \Phi_1 = 0.5\pi $). 
    Equal Rashba and Dresselhaus~parameters are chosen since the interference of the Rashba and Dresselhaus spin-orbit~fields leads then to the effective spin-orbit~fields
    \begin{align}
        \boldsymbol{\Omega}_\mathbf{L} &= [0 , \, -2\alpha_\mathrm{L} k_x , \, 0] 
        \intertext{and}
        \boldsymbol{\Omega}_\mathbf{R} &= -[0 , \, -2\alpha_\mathrm{R} k_x , \, 0] .
    \end{align}
    Tunneling~electrons are therefore subject to the maximal SOCs if the magnetizations are aligned along~$ \mp \hat{x} $~(due to the $ k_x $~dependence) and do not at all experience any SOC for magnetizations parallel to~$ \mp \hat{y} $~(as the $ k_y $~dependence dropped out), eventually raising the maximally possible magnetoanisotropy.

    As a result, the additionally generated unconventional Andreev-reflection conductance~contribution becomes maximal at~$ \Phi_1 = 0 $ and completely vanishes at~$ \Phi_1 = 0.5 \pi $, explaining the overall substantial conductance~decrease when increasing $ \Phi_1 $ from~$ \Phi_1 = 0 $ to~$ \Phi_1 = 0.5 \pi $. 
    Recall that conventional Andreev~reflections are suppressed in junctions with ``thin'' superconducting links and the tunneling~conductance is, apart from partially through SOCs allowed unconventional Andreev~reflections, largely determined by electron~transmissions. 
    Surprisingly, and in sharp contrast to two-electrode F/S~junctions in which they even fully disappear there~\cite{Hoegl2015,*Hoegl2015a}, unconventional Andreev~reflections~(at~$ \Phi_1 = 0 $) are most likely at energies around the gap~edge~($ eV \approx \Delta_0 $), facilitating a somewhat broadened conductance~peak~(``conductance~shoulder''). 
    As before, the tunneling~conductance at energies well above~$ \Delta_0 $ mostly stems from electron~transmissions, and just slightly decreases with increasing~$ \Phi_1 $ owing to the effectively slightly lowered tunneling~probability.

    \begin{figure}
        \centering
        \includegraphics[width=0.45\textwidth]{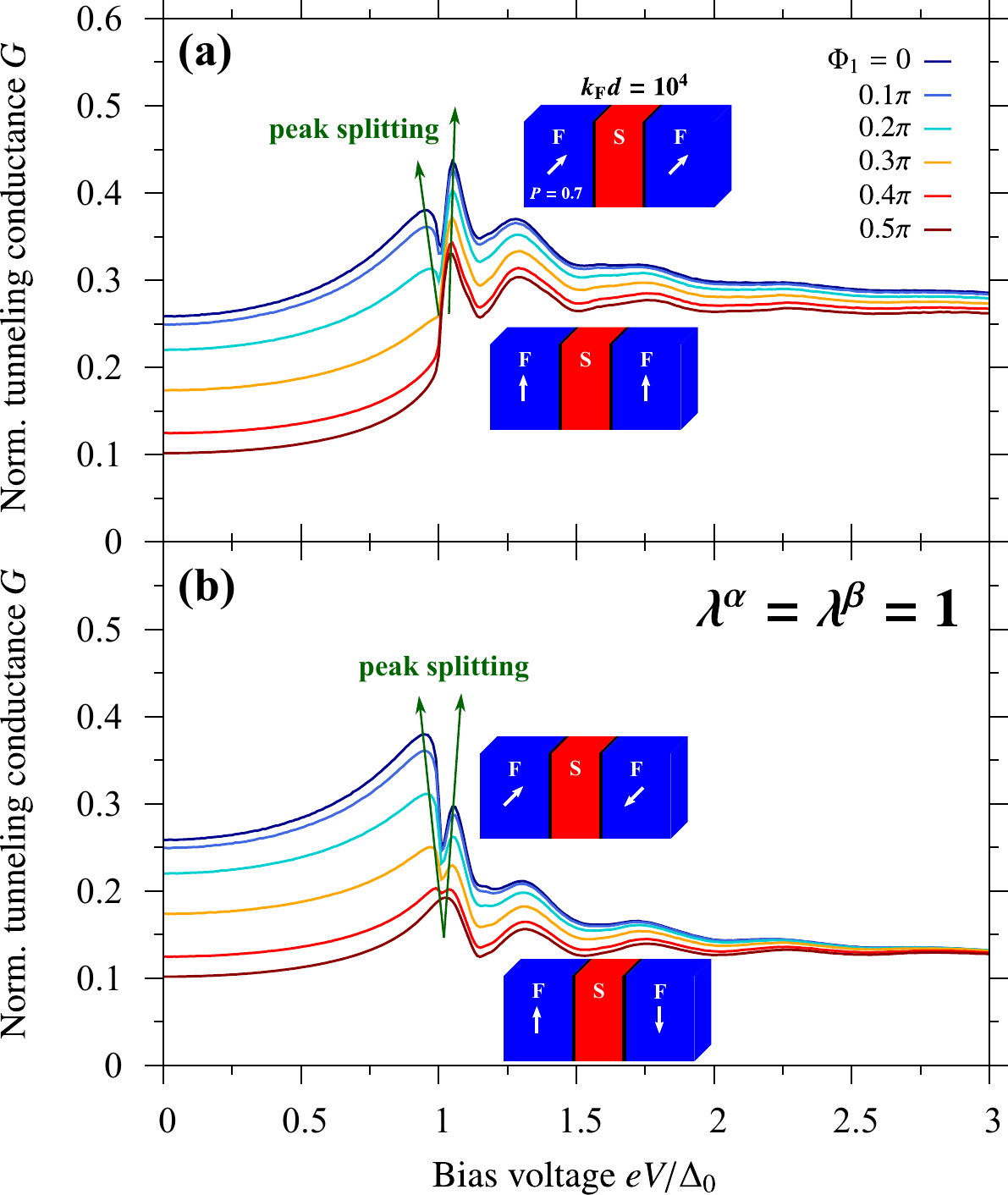}
        \caption{Same calculations as in~Fig.~\ref{Fig_GeneralConductanceWithSOC_Short_MagnetizationTriplet}, but assuming a \emph{``thick'' superconducting link} of thickness~$ d = 10^4 / k_\mathrm{F} $. 
            The gap-edge conductance-peak~splitting serves again as a precursor of induced superconducting triplet~pairings. 
        }
        \label{Fig_GeneralConductanceWithSOC_Long_MagnetizationTriplet}
    \end{figure}

    Analogously, we present the tunneling~conductance's magnetization-angle~dependence for a F/S/F~junction containing a ``thick'' superconducting link~($ d = 10^4 / k_\mathrm{F} $) in~Fig.~\ref{Fig_GeneralConductanceWithSOC_Long_MagnetizationTriplet}; all other parameters are not changed. 
    As we argued above, the unconventional Andreev-reflection conductance~contribution gets maximal at~$ \Phi_1 = 0 $ and vanishes at~$ \Phi_1 = 0.5\pi $. 
    Besides remarkably enhancing the subgap tunneling~conductance, the superconducting triplet~pairings induced by unconventional Andreev~reflections split the gap-edge conductance~peak again into two distinct peaks. 
    Just as we encountered when analyzing this feature earlier, the peak~splitting becomes most pronounced when simultaneously the unconventional Andreev-reflection~process dominates the subgap tunneling~conductance, i.e., at~$ \Phi_1 = 0 $ in our case. 
    Note that the split conductance~peaks' amplitude~ratios in the parallel and antiparallel magnetization~configurations are opposite, as one might also observe in~Fig.~\ref{Fig_GeneralConductanceWithSOC_Long}(b). 
    While the peak slightly above the gap corresponds to the larger tunneling~conductance for parallel magnetizations---as the tunneling~conductance gets there amplified by additionally allowed electron~transmissions---it is the peak below the gap that entails maximal tunneling~conductances in the antiparallel magnetized scenario. 
    Finally, the aforementioned geometrical~oscillations at~$ eV > \Delta_0 $ are clearly visible and not notably altered by rotating the magnetization~orientations, while the conductance~amplitudes decrease slightly with increasing~$ \Phi_1 $ there due to the slightly reduced interfacial transparencies, just as we explained for ``thin'' superconducting links above.

    While Andreev~reflections are only barely impacted by switching the ferromagnets' \emph{relative} magnetizations from the parallel to the antiparallel orientations~(recall our discussions in~Sec.~\ref{Sec_General}), the last paragraphs demonstrated that they are nonetheless extremely sensitive to rotations of the ferromagnets' \emph{absolute} magnetization~directions in the presence of interfacial SOCs, and marked magnetoanisotropies in the tunneling~conductance can occur. 
    To quantify the latter, and emphasize that they predominantly originate from the strongly magnetoanisotropic (unconventional) Andreev-reflection~probabilities, an earlier work on two-electrode F/S~junctions established the \emph{in-plane magnetoanisotropic Andreev~reflection~(MAAR)}~\cite{Hoegl2015,*Hoegl2015a} 
    \begin{equation}
        \mathrm{MAAR} (\Phi_1) = \frac{G(0) - G(\Phi_1)}{G(\Phi_1)} .
    \end{equation}
    One could evaluate the MAAR, for~instance, deep inside the superconducting junction~regime at~$ eV=0 $, and compare the values against its normal-state~counterpart at~$ eV \gg \Delta_0 $. 
    As we can already expect from the large  magnetization-controlled tunability of the absolute conductance~amplitudes in~Figs.~\ref{Fig_GeneralConductanceWithSOC_Short_MagnetizationTriplet} and~\ref{Fig_GeneralConductanceWithSOC_Long_MagnetizationTriplet}, unconventional Andreev~reflections could entail huge superconducting MAAR~ratios, which can easily exceed the equivalent normal-state tunneling~anisotropic~magnetoresistance~\cite{Moser2007,MatosAbiague2009} by more than three orders of magnitude in the half-metallic~limit and can be further enhanced by increasing the thickness of the superconducting~link~$ d $. 
    While the calculated MAAR in a junction with a ``thin'' superconducting link~($ d = 10^3 / k_\mathrm{F} $) lies clearly below the values predicted for comparable two-electrode F/S~junctions~(as Andreev~reflection is not the dominant scattering~process there), ``thick'' superconducting links~($ d = 10^4 / k_\mathrm{F} $) cause MAARs that already remarkably overcome those in the corresponding F/S~junctions~(as Andreev~reflection dominates now the subgap~regime).   
    As the overall physics and qualitative characteristics are basically similar to the F/S~case, we do not deeply analyze our MAAR~calculations here.

    \section{Magnetoresistance effects    \label{Sec_TMR}}

    MR~effects count to the probably most intensively investigated phenomena in magnetic spin-valve~junctions. 
    Our work offers the possibility to study the MR of \emph{superconducting} magnetic spin~valves, and elaborate more on the ramifications of the competition between the usual---and in normal-state spin~valves dominant---electron~transmissions and the superconducting~junctions' unique Andreev~reflections. 
    Adapting its most common definition, the MR~ratio at an \emph{absolute} magnetization~orientation determined by~$ \Phi_1 $~(in the left ferromagnet) is given by
    \begin{equation}
        \mathrm{MR (\Phi_1) } = \frac{G_\mathrm{P} - G_\mathrm{AP}}{G_\mathrm{AP} } = \frac{G(\Phi_1) - G(\Phi_1 + \pi) }{ G(\Phi_1 + \pi) } ,
        \label{Eq_TMR}
    \end{equation}
    where~$ G_\mathrm{P} = G(\Phi_1) $~[$ G_\mathrm{AP} = G(\Phi_1 + \pi) $] indicates the tunneling~conductance in the parallel~(antiparallel) magnetization~configurations and can be extracted from~Eq.~\eqref{Eq_Conductance} at zero temperature.

    \begin{figure}
        \centering
        \includegraphics[width=0.45\textwidth]{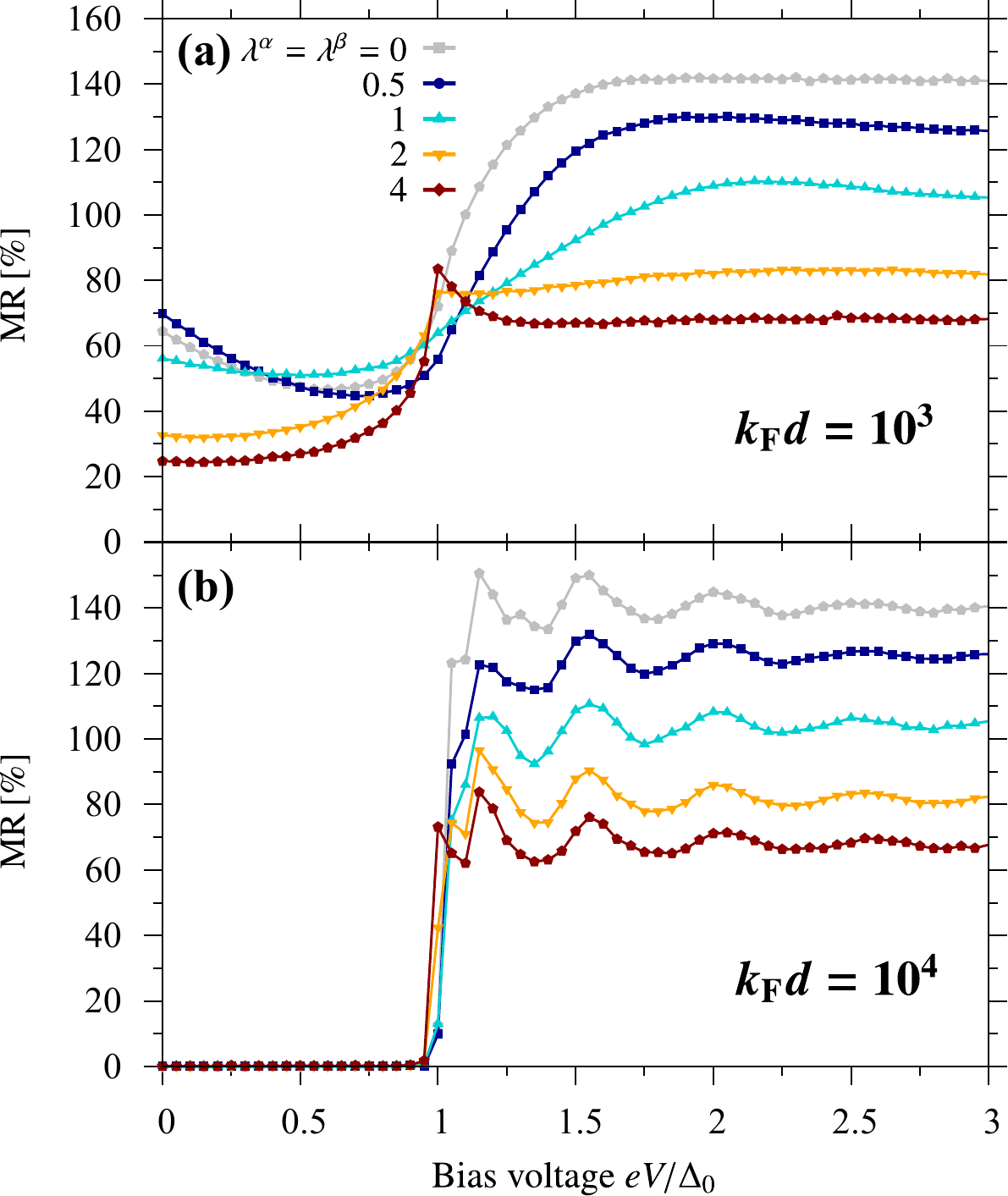}
        \caption{Calculated MR as a function of the applied bias~voltage~$ V $ and for different indicated uniform SOC~strengths~$ \lambda^\alpha = \lambda^\beta $~($ = \lambda^\alpha_\mathrm{L} = \lambda^\alpha_\mathrm{R} = \lambda^\beta_\mathrm{L} = \lambda^\beta_\mathrm{R} $), assuming a (a)~\emph{``thin''} and (b)~\emph{``thick'' superconducting link} of thicknesses~$ d = 10^3 / k_\mathrm{F} $ and $ d = 10^4 / k_\mathrm{F} $, respectively. 
            The spin~polarization of the ferromagnets is~$ P= 0.7 $ and the magnetizations are aligned along $ \mp \hat{y} $.  
        }
        \label{Fig_TMR_VS_BiasVoltage}
    \end{figure}

    In~Fig.~\ref{Fig_TMR_VS_BiasVoltage}, we illustrate the computed MR~[more precisely, $ \mathrm{MR}(\Phi_1 = \pi / 2) $] as a function of the applied bias~voltage~$ V $ for various strengths of interfacial Rashba and Dresselhaus~SOCs~$ \lambda^\alpha_\mathrm{L} = \lambda^\alpha_\mathrm{R} = \lambda^\beta_\mathrm{L} = \lambda^\beta_\mathrm{R} $, distinguishing again between junctions with ``thin''~($ d = 10^3 / k_\mathrm{F} $) and ``thick''~($ d = 10^4 / k_\mathrm{F} $) superconducting links. 
    As expected from our earlier analyses of the conductance~features, the MR~ratios always reach their maximal values whenever the underlying tunneling~conductance is dominated by electron~transmissions, i.e., at $ eV \geq \Delta_0 $. 
    In the subgap bias-voltage~regime, even small conductance~contributions originating from conventional and---in the presence of nonzero interfacial SOCs---unconventional Andreev~reflections immediately lower the resulting MR. 
    This observation explains the MR~suppression with increasing SOC~strength in the junction with the ``thin'' link, as well as the fully vanishing subgap MR for the ``thick'' link. 
    In the first case~(``thin'' link), increasing the SOC~parameters above~$ \lambda^\alpha_\mathrm{L} = \lambda^\alpha_\mathrm{R} = \lambda^\beta_\mathrm{L} = \lambda^\beta_\mathrm{R} \approx 0.5 $ raises a marked conductance~enhancement owing to unconventional Andreev~reflections. 
    Since Andreev~reflections are much less sensitive to changes of the \emph{relative} magnetization~orientations than the in the absence of SOCs dominant electron~transmissions, the related MR starts to be remarkably damped, finally resulting in a complete suppression if the conductance is exclusively determined by (unconventional) Andreev~reflections as we witness in the second case~(``thick'' link). 
    At voltages above the gap~($ eV > \Delta_0 $), the MR mostly reveals the conductance~properties resulting from usual electron~transmissions, including its monotonic decrease with increasing SOC~strengths; the interfacial SOCs act then similarly to additional interfacial barriers that suppress electron~transmissions and thus the MR. 
    Interestingly, the geometrical conductance~oscillations caused by coherent electron~transmissions through ``thick'' superconducting~links are moreover transferred into the respective MR--bias~voltage characteristics. 
    Note that the MRs in the junctions' normal-state~counterparts~(recovered at~$ eV \gg \Delta_0 $) are---analogously to the related tunneling~conductances~(compare, e.g., Fig.~\ref{Fig_GeneralConductanceWithSOC_Short} to Fig.~\ref{Fig_GeneralConductanceWithSOC_Long} supposing~$ eV \gg \Delta_0 $)---nearly completely independent of the link~thickness, which is a consequence of our fully ballistic description. 
    Regarding the maximally possible MR~amplitudes, half-metallic~junctions~(with spin~polarizations~$ P \to 1 $) are certainly the most auspicious candidates; similarly to normal-conducting~systems, maximal MRs easily reach then values above~$ 1000 \, \% $.

    Summarizing the preceding paragraphs, the MR~features~(amplitudes) are predominantly controlled by the intriguing competition between Andreev~reflections~(dominant in the subgap~regime, $ eV < \Delta_0 $) and electron~transmissions~(dominant at~$ eV > \Delta_0 $). 
    Most relevant to future experimental studies might therefore be exploring the MR exactly at the gap-edge~energy, i.e., at~$ eV = \Delta_0 $, at which the aforementioned competition between Andreev~reflections and electron~transmissions becomes most pronounced. 
    We will address the experimental signatures in~Sec.~\ref{Sec_ATMR}.

    \section{Anisotropic magnetoresistance    \label{Sec_ATMR}}

\begin{figure}
        \centering
        \includegraphics[width=0.49\textwidth]{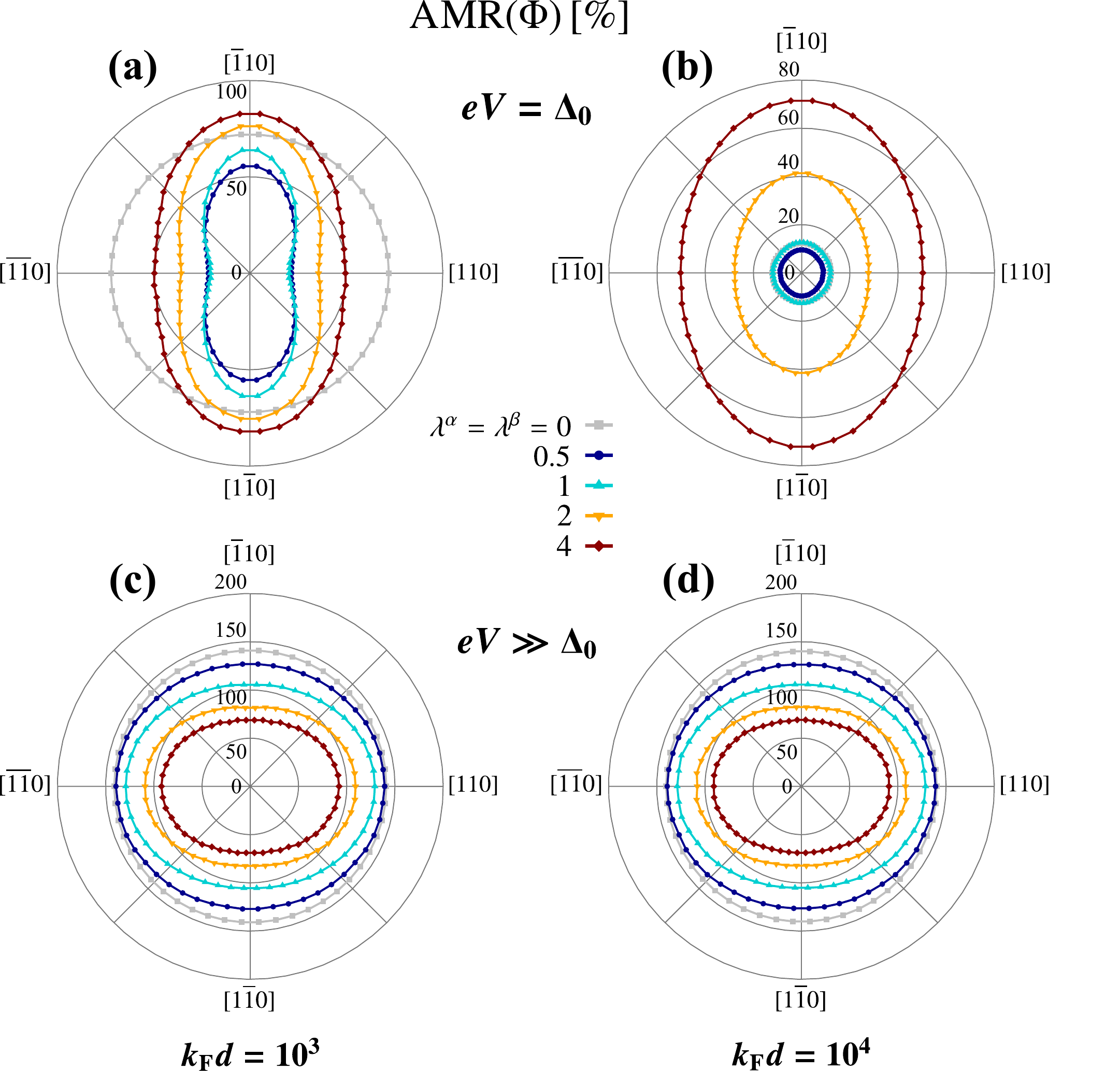}
        \caption{Calculated AMR~(i.e., the angular dependence of the MR on the left ferromagnet's magnetization~orientation~$ \Phi_1 = \Phi $) for different indicated uniform SOC~strengths~$ \lambda^\alpha = \lambda^\beta $~($ = \lambda^\alpha_\mathrm{L} = \lambda^\alpha_\mathrm{R} = \lambda^\beta_\mathrm{L} = \lambda^\beta_\mathrm{R} $), assuming a (a)~\emph{``thin''} and (b)~\emph{``thick'' superconducting link}~(thicknesses~$ d = 10^3 / k_\mathrm{F} $ and $ d = 10^4 / k_\mathrm{F} $), and setting~$ eV = \Delta_0 $. 
            The remaining parameters are the same as in~Fig.~\ref{Fig_TMR_VS_BiasVoltage}. 
            For~comparison, panels~(c) and (d) show the corresponding \emph{normal-state} AMR~ratios. 
        }
        \label{Fig_ATMR}
    \end{figure}

    As we thoroughly discussed in~Sec.~\ref{Sec_MAAR}, interfacial SOCs usually entail marked magnetoanisotropies in experimentally probeable transport~quantities. 
    Consequently, not only the tunneling~conductance itself, but also closely related measures, like the MR, strongly depend on the \emph{absolute} orientation of the ferromagnets' magnetization~directions. 
    We emphasized that in~Eq.~\eqref{Eq_TMR} through explicitly stating the MR's $ \Phi_1 $~dependence. 
    While the magnetoanisotropic MR---usually referred to as \emph{anisotropic magnetoresistance~(AMR)}---has already been comprehensively analyzed in normal-conducting F/N/F~junctions~\cite{MatosAbiague2009}, characterizations of AMR~phenomena in superconducting junctions have yet been missing. 
    To close this gap, we present the angular dependence of the considered F/S/F~junction's AMR~[note that $ \mathrm{AMR} (\Phi) = \mathrm{MR} (\Phi_1 = \Phi) $; recall~Eq.~\eqref{Eq_TMR}] for various uniform SOC~strengths in~Fig.~\ref{Fig_ATMR}. 
    Motivated by our previous arguments, we focus on the bias~voltage~$ eV = \Delta_0 $, at which we expect a strong competition between Andreev~reflections and electron~transmissions, to compare the results in the superconducting to those in the normal-conducting~($ eV \gg \Delta_0 $) scenario; the thickness of the ``thin'' link is again~$ d = 10^3 / k_\mathrm{F} $ and that of the ``thick'' link~$ d = 10^4 / k_\mathrm{F} $.

    As long as interfacial SOCs are completely absent, the (A)MR is isotropic, and the MR~amplitudes do hence not alter as the \emph{absolute} direction of the ferromagnets' magnetizations gets rotated~(gray circles). 
    Already weak SOCs, however, notably tilt the AMR~curves~(more elliptical, colored, curves), and give rise to clearly magnetoanisotropic MRs, with substantially larger MR~ratios at magnetizations pointing along $ \mp \hat{y} $~(as unconventional Andreev~reflections get suppressed there and electron~transmissions dominate). 
    Although qualitatively similar physics occurs in the normal~state, the ``tilting''~(which is directly linked to the ``strength'' of the MR~magnetoanisotropy, as we will elaborate on later) is much weaker than in the superconducting case~(whereas the overall MR~values are more than twice as large as in the superconducting junction due to the dominant electron~transmissions in normal-state~junctions) and it is now the magnetization along~$ \mp \hat{x} $ that results in (slightly) larger MR~amplitudes.

    These observations suggest that, while overall large MR~ratios indicate dominant electron~transmissions, marked AMR magnetoanisotropies serve as an experimentally accessible signature of dominant Andreev~reflections. 
    These contrary features could be beneficial to subsequent experimental works to disentangle Andreev-reflection- from electron-transmission-related physics. 
    As they both predominantly originate from the peculiar Andreev-reflection~process, MAAR and AMR~effects share all their fundamental properties.  
    For~instance, and similarly to the aforementioned in-plane MAAR, it is vital to the AMR that interfacial Rashba \emph{and} Dresselhaus SOCs interfere. 
    If just Rashba or Dresselhaus~SOCs alone were present, the (A)MR would become fully isotropic. 
    Maximally anisotropic AMRs arise again if the Rashba and Dresselhaus SOC~strengths are equal, as we likewise explained in connection with the tunneling-conductance magnetoanisotropies in~Sec.~\ref{Sec_MAAR}.

    To quantify the ``strength'' of the MR~magnetoanisotropy, and relate it to the ``tilting'' of the AMR~curves visible in~Fig.~\ref{Fig_ATMR}, we introduce the \emph{AMR~efficiency}~\cite{MatosAbiague2009} 
    \begin{equation}
        \eta = \frac{\mathrm{AMR} (\Phi = \pi/2) - \mathrm{AMR} (\Phi=0)}{\mathrm{AMR} (\Phi=0)} ,
        \label{Eq_ATMREff}
    \end{equation}
    which essentially measures the relative change of the MR~ratios while the ferromagnets' magnetizations are rotated from the $ \mp \hat{x} $- toward the $ \mp \hat{y} $-orientation.

    \begin{figure*}
        \centering
        \includegraphics[width=0.95\textwidth]{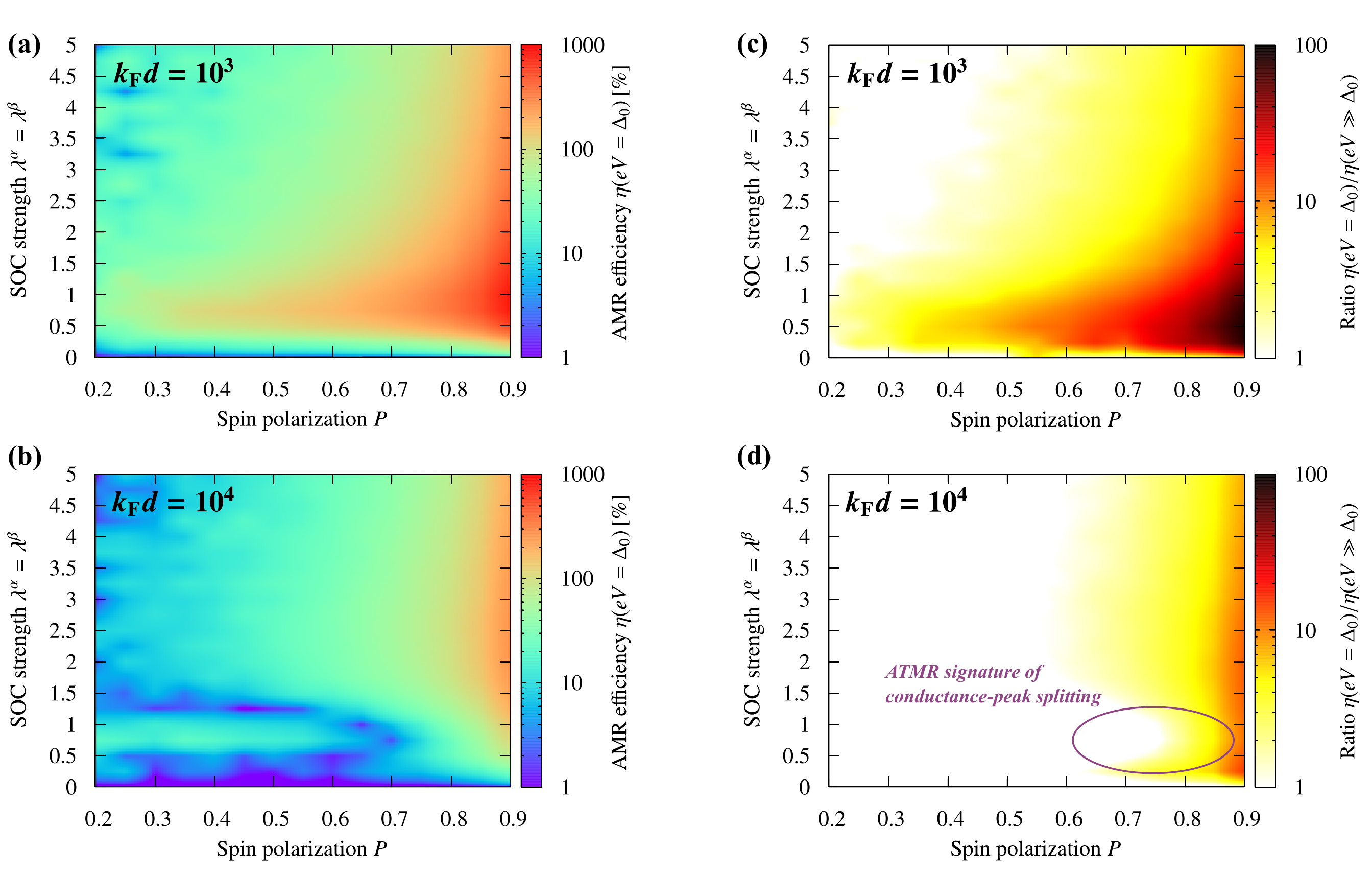}
        \caption{Calculated gap-edge AMR~efficiency~$ \eta (eV = \Delta_0) $ as a function of the ferromagnets' spin~polarization~$ P $ and the uniform SOC~strength~$ \lambda^\alpha = \lambda^\beta $~($ = \lambda^\alpha_\mathrm{L} = \lambda^\alpha_\mathrm{R} = \lambda^\beta_\mathrm{L} = \lambda^\beta_\mathrm{R} $) in the presence of a (a)~\emph{``thin''} and (b)~\emph{``thick'' superconducting link}~(thicknesses $ d = 10^3 / k_\mathrm{F} $ and $ d = 10^4 / k_\mathrm{F} $). 
            Unconventional Andreev~reflections can substantially enhance the AMR~efficiency in the superconducting state~(up to about two orders of magnitude), as inspecting the ratios between the gap-edge AMR~efficiency and its normal-state~counterpart in~(c) and~(d) illustrates. 
            The locally suppressed AMR~efficiency~(violet) in~(d) signifies the gap-edge conductance-peak splitting that we attributed to superconducting triplet~pairings~(see~Sec.~\ref{Sec_General}). 
        }
        \label{Fig_ATMR_ColorMap}
    \end{figure*}

    Figure~\ref{Fig_ATMR_ColorMap} shows the computed AMR~efficiencies as functions of the ferromagnets' spin~polarization~$ P $ and the uniform SOC~strengths~$ \lambda^\alpha_\mathrm{L} = \lambda^\alpha_\mathrm{R} = \lambda^\beta_\mathrm{L} = \lambda^\beta_\mathrm{R} $, as well as the ratio between the AMR~efficiencies in the superconducting and normal-conducting states, respectively. 
    The results further substantiate our previous claims, and demonstrate the peculiar role of unconventional Andreev~reflections for another time. 
    More specifically, the MR~magnetoanisotropies of junctions containing the ``thin'' superconducting link become most pronounced~(resulting in the largest AMR~efficiencies~$ \eta $) as the ferromagnets' spin~polarization approaches the half-metallic~limit~($ P \to 1 $) and the SOC~strengths are tuned to~$ \lambda^\alpha_\mathrm{L} = \lambda^\alpha_\mathrm{R} = \lambda^\beta_\mathrm{L} = \lambda^\beta_\mathrm{R} \approx 1 $. 
    As we pointed out in~Sec.~\ref{Sec_General}, those parameters maximize the unconventional Andreev-reflection~contribution to the tunneling~conductances~(at the considered bias~voltage~$ eV = \Delta_0 $), which responds most sensitively to changes of the \emph{absolute} magnetization~directions and entails huge MR~magnetoanisotropies~($ \eta $). 
    Noteworthy, the AMR~efficiency in the superconducting state exceeds its normal-state~counterpart by more than two orders of magnitude.

    The arguments provided in the preceding paragraph do, in~principle, also hold for junctions with the ``thick'' superconducting link. 
    Nevertheless, the MR~magnetoanisotropies~(amplitudes of $ \eta $) appear to be substantially lower in that case, which must indicate that unconventional Andreev~reflections are additionally suppressed. We indeed observe this suppression most clearly at spin~polarization~$ P=0.7 $ and SOC~parameters~$ \lambda^\alpha_\mathrm{L} = \lambda^\alpha_\mathrm{R} = \lambda^\beta_\mathrm{L} = \lambda^\beta_\mathrm{R} \approx 1 $, for which we analyzed the underlying tunneling~conductances in all details in~Sec.~\ref{Sec_General} unraveling the conductance-peak splitting at~$ eV \approx \Delta_0 $ as a transport~signature of superconducting triplet~pairings. 
    As a result, the initially present conductance~peak exactly at~$ eV = \Delta_0 $ turns into a conductance~dip, the unconventional Andreev-reflection~contribution gets damped, and the calculated MR~magnetoanisotropy~($ \eta $) must therefore notably drop. 
    The huge MR~magnetoanisotropies~($ \eta $) would instead occur slightly below and above the superconducting gap, corresponding to the two newly forming conductance double~peaks. 
    The latter can hence not only be identified in the tunneling-conductance~data, but leave also an indirect imprint on the AMR~characteristics. 
    Though the unconventional Andreev-reflection~contribution at~$ eV = \Delta_0 $ is small(er), it is large enough to raise AMR~efficiencies that still overcome their normal-state~counterparts by more than one order of magnitude.

    \section{Conclusions    \label{Sec_Conclusions}}

    To summarize, we studied the tunneling-conductance~features of superconducting magnetic F/S/F spin-valve junctions paying special attention to the ramifications of interfacial Rashba and Dresselhaus~SOCs. 
    We distinguished between junctions hosting ``thin''~(thickness of about $ 125 \, \mathrm{nm} $) and ``thick''~(thickness of about $ 1.25 \, \upmu \mathrm{m} $) superconducting links, and allowed for arbitrary in-plane orientations of the magnetization~directions inside the ferromagnetic electrodes. 
    Interfacial SOCs facilitate unconventional (spin-flip) Andreev~reflections at junction~interfaces that are commonly expected to be at the heart of numerous transport~anomalies, as they concurrently introduce spin-polarized superconducting triplet~pairings into the system. 
    Regarding the considered F/S/F~junctions, we observed that unconventional Andreev~reflections can give rise to a conductance-peak~splitting close to the singlet-gap~energy, which eventually reveals the interplay between the usual superconducting singlet and the additional effectively induced triplet gaps. 
    Owing to their close connection to unconventional Andreev~reflections, we demonstrated that these peak~splittings---and at the same time also the overall amplitudes of the tunneling~conductance---are efficiently tunable through altering the ferromagnets' \emph{absolute} magnetization~orientations. 
    We eventually quantified the MR of superconducting spin-valve junctions, and unraveled that unconventional Andreev~reflections~(and thus indirectly the present SOCs) furthermore lead to marked MR~magnetoanisotropies, which we termed AMRs. 
    Measuring the AMR~efficiency~(i.e., the ``strength'' of the MR~magnetoanisotropy) provides another experimental possibility to detect the  triplet-pairing signifying gap-edge conductance double~peaks. 
    In view of future experiments, we suggest focusing on highly spin-polarized junctions, in which the strong spin~filtering of transmitted electrons yields overall giant MRs~(analogously to Julli\`{e}re's model), but which still entail a considerable amount of unconventional Andreev~reflections to simultaneously raise huge magnetoanisotropies~(MAARs and AMRs).

    \begin{acknowledgments}
        This work was supported by Deutsche Forschungsgemeinschaft~(DFG, German Research Foundation) through Subproject~B07 within the Collaborative~Research~Center SFB~1277~(Project-ID~314695032) and the Research~Grant ``Spin and magnetic properties of superconducting tunnel~junctions''~(Project-ID~454646522). 
    \end{acknowledgments}

    \bibliography{paper}

\end{document}